\documentclass[10pt,twocolumn,letterpaper]{article}
\usepackage[pagenumbers]{cvpr} % For arXiv versions

\usepackage{multirow}

\usepackage{enumitem}
\usepackage{graphicx}
\usepackage{xspace}
\usepackage{mathtools}
\usepackage{threeparttable}
\usepackage{array}

\usepackage{tikz}
\usetikzlibrary{positioning,fit,arrows.meta,patterns}
\usepackage{adjustbox}
\usepackage{pgfplots}
\pgfplotsset{compat=1.17}

\usepgfplotslibrary{groupplots}
\usepackage{placeins}
\usepackage{makecell}
\usepackage{color}
\usepackage{float}
\usepackage{algorithm}
\usepackage{algpseudocode}
\usepackage{tcolorbox}

\usepackage{amsmath,amssymb,amsfonts}

\usepackage{tabularx,array,booktabs}
\usepackage{colortbl}

\definecolor{systemcolor}{HTML}{E8F1FF}
\definecolor{usercolor}{HTML}{F7FFE9}
\definecolor{assistantcolor}{HTML}{FFF7E9}

\newcommand{\compw}{0.4\linewidth}
\newcommand{\hpw}{0.70\linewidth}

\definecolor{cvprblue}{rgb}{0.21,0.49,0.74}
\usepackage[pagebackref,breaklinks,colorlinks,allcolors=cvprblue]{hyperref}

\def\paperID{21048} % * Enter the Paper ID here
\def\confName{CVPR}
\def\confYear{2026}

\title{M3DocDep: Multi-modal, Multi-page, Multi-document Dependency Chunking with Large Vision-Language Models}

\author{Joongmin Shin$^{1}$ \quad Jeongbae Park$^{1}$ \quad Jaehyung Seo$^{2\ddagger}$ \quad Heuiseok Lim$^{1,3\ddagger}$\\
$^{1}$Human-inspired AI Research, Korea University\\
$^{2}$Computer Science and Engineering, Konkuk University\\
$^{3}$Department of Computer Science and Engineering, Korea University\\
{\tt\small \{tlswndals13, insmile, limhseok\}@korea.ac.kr \quad seojae777@konkuk.ac.kr}\\
{\small $^{\ddagger}$Corresponding authors}
}

\newcommand{\ours}{\textsc{M3DocDep}\xspace}

\begin{document}
\maketitle

\begin{abstract}
In long, multi-page industrial documents, retrieval-augmented generation (RAG) depends heavily on whether chunk boundaries follow the document's true structure. Existing text-centric chunkers and generative hierarchy parsers often miss cross-page parent--child relations, figure/table--caption bindings, and boundary cues, which leads to fragmented or redundant chunks and degrades both retrieval and answer quality. We propose \textbf{\ours}, an LVLM-based pipeline that first recovers block-level dependencies and then constructs chunks along the recovered document tree. The pipeline uses SharedDet as a common DP+OCR preprocessing layer, extracts multi-modal block embeddings with boundary-aware SoftROI pooling, scores candidate parent--child edges with a biaffine head, decodes a globally valid dependency tree with MST constraints, and builds tree-guided chunks annotated with section paths and page ranges. Under a shared-block evaluation protocol, \ours improves STEDS by +28.5--39.6\% on DHP benchmarks, retrieval nDCG by +1.1--15.3\%, and QA ANLS by +4.5--15.3\% on corpus-level RAG benchmarks. These results show that recovering document dependencies before chunking yields more coherent retrieval units for long, multi-page multi-modal documents.

\end{abstract}

\section{Introduction}
Retrieval-augmented generation (RAG) has become a core mechanism for enabling large language models (LLMs) to handle long, information-dense documents~\cite{lewis2021retrievalaugmented, Jeong2023ASO, Ge2023Development}. However, its effectiveness critically depends on how documents are chunked into semantically coherent units, the retrieval granularity that governs both retrieval precision and answer accuracy~\cite{gao2024retrievalaugmentedgenerationlargelanguage}. Yet prevailing text-centric chunkers overlook visual and structural cues in real-world documents, making them brittle on scanned pages, multi-page PDFs, and complex industrial layouts~\cite{Gong2020Recurrent, qu-etal-2025-semantic}. This issue is amplified by OCR noise and misalignment, which cause duplicated or ambiguous chunks and ultimately degrade retrieval and QA performance~\cite{tito2023hierarchicalmultimodaltransformersmultipage, hong-etal-2024-intelligent}. 

Structure-aware chunking based on vision-driven Document Parsing (DP)~\cite{dosovitskiy2021vit,10.1145/3534678.3539043} can robustly extract visually coherent regions such as tables and text blocks~\cite{yepes2024financialreportchunkingeffective}. However, it still fails to capture the semantic hierarchies of multi-page documents (e.g., parent--child relations), leaving full-document context insufficiently modeled~\cite{xing2024dochienet,hong-etal-2024-intelligent}. To address this, recent approaches combine DP, OCR, and LLMs to learn or infer Document Hierarchical Parsing (DHP) via instruction fine-tuning (SFT)~\cite{zhang2024instructiontuninglargelanguage}, enabling chunking to follow document structure~\cite{rausch2021docparser,rausch2023dsgendtoenddocumentstructure,dochienet,multidocfusion}. However, converting pages into pure text for LLM input inevitably removes crucial visual cues such as color, font size, and typographic emphasis, which hinders accurate hierarchy reconstruction and limits the handling of figures and tables~\cite{zhang-etal-2024-llm-graph,wang-etal-2024-docllm,tabatabaei-etal-2025-large}.

In contrast, large vision--language models (LVLMs) pretrained at scale can jointly interpret visual and textual content, making them well suited to the multi-modal signals found in diverse industrial documents~\cite{wang2025internvl35advancingopensourcemultimodal,an2025llavaonevision15fullyopenframework,bai2025qwen25vltechnicalreport}. However, SFT-based LVLM approaches still often struggle to recover a globally consistent hierarchy over long multi-page documents: cross-page references are unstable, visual cues are only partially preserved after textualization, and sequence generation does not naturally enforce tree constraints~\cite{xing-etal-2025-intelligent}. This leads to a recurring failure mode in document RAG: if block dependencies are recovered inaccurately, chunk boundaries also become unreliable, which in turn harms retrieval precision and answer grounding.

To address this problem, we introduce \textbf{\ours}, an LVLM-based dependency chunking pipeline that explicitly follows the chain
\emph{better block dependency recovery $\rightarrow$ better document tree $\rightarrow$ better chunk boundaries $\rightarrow$ better retrieval and QA}.
\ours first constructs a shared block canvas with SharedDet (DP+OCR), then extracts multi-modal block embeddings, scores candidate parent--child edges, decodes a Global Document Dependency Tree, and finally builds tree-guided chunks aligned with section structure and figure/table--caption relations. By recovering dependencies before chunking, \ours produces more coherent retrieval units that preserve long-range cross-page structure while remaining compatible with corpus-level RAG.

\paragraph{Contributions}
\begin{itemize}[leftmargin=*]
\item We introduce an LVLM-based dependency scoring framework that reconstructs a global document tree from multi-modal block representations, including long-range cross-page parent--child relations.

\item We propose tree-guided structure-aware chunking that follows recovered section subtrees, preserves figure/table--caption bindings, and annotates each chunk with section-path metadata for corpus-level retrieval.

\item Under a shared evaluation protocol---the same SharedDet blocks, the same chunk budget, and the same retrievers/readers across methods---\textbf{\ours} improves hierarchy recovery, retrieval quality, and QA accuracy across diverse industrial document corpora, while remaining effective under DP/OCR/embedding swaps tested in this work.
\end{itemize}

\section{Related Work}

\paragraph{Chunking for QA on Long Industrial Documents}
Chunking is essential for handling long, multi-page documents in RAG~\cite{gao2024retrievalaugmentedgenerationlargelanguage}. Early approaches rely on length-based~\cite{Gong2020Recurrent} or semantic chunking~\cite{qu-etal-2025-semantic}, but they fail to capture document hierarchies or integrate visual layout elements such as tables and figures. LLM-based methods (e.g., LumberChunker~\cite{duarte2024lumberchunker}, Perplexity chunking~\cite{zhao2024metachunking}) improve semantic grouping but still suffer from fragmentation because they do not explicitly model hierarchical structure or visual cues. StyleDFS~\cite{hong-etal-2024-intelligent} highlights the importance of hierarchical analysis but degrades on scanned or irregular documents. MultiDocFusion~\cite{multidocfusion} combines DP, OCR, and LLMs to infer hierarchies and improve chunking, but it remains limited by LLM context windows and loses critical visual cues during textualization. Recent Multi-modal LVLMs~\cite{wang2025internvl35advancingopensourcemultimodal,an2025llavaonevision15fullyopenframework,bai2025qwen25vltechnicalreport} excel at image-centric tasks but cannot process full multi-page documents due to context window constraints. These limitations highlight the need for chunking methods that jointly capture visual layout, hierarchical structure, and cross-page context~\cite{saadfalcon2023pdftriagequestionansweringlong,10.1007/978-3-031-70552-6_13}. Compared with prior structure-then-chunk pipelines such as MultiDocFusion, our method recovers hierarchy through multimodal block embeddings and globally constrained MST decoding rather than autoregressive hierarchy generation. This design preserves figure/table regions together with captions inside a recovered tree and is complementary to retriever-side multimodal RAG methods that can operate on top of the chunks produced here.

\paragraph{Document Parsing and Document Hierarchical Parsing}
Document parsing (DP) methods segment PDFs and scanned documents into visual components such as tables, figures, and text blocks~\cite{dosovitskiy2021vit,10.1145/3534678.3539043}, but their detection-centric design cannot recover semantic hierarchies (e.g., ``1.2--1.2.1''), losing global context~\cite{rausch2023dsgendtoenddocumentstructure,wang-etal-2024-docllm}. Document Hierarchical Parsing (DHP) approaches~\cite{rausch2021docparser,rausch2023dsgendtoenddocumentstructure,zhang-etal-2024-pdf} attempt to restore such structure but suffer from template bias, poor generalization to scanned or irregular layouts, and difficulty with multi-page dependencies~\cite{WANG2024110836,xing2024dochienet}. LLM-based methods provide long-context reasoning~\cite{fujitake-2024-layoutllm} but underutilize visual cues and remain context-limited, leading to incomplete hierarchy reconstruction~\cite{zhang-etal-2024-llm-graph,wang-etal-2024-docllm,tabatabaei-etal-2025-large}. Furthermore, decoder-style SFT is misaligned with tree constraints and often recovers only partial header hierarchies while relying on rules for the rest~\cite{multidocfusion}. LVLMs jointly process visual and textual signals, but SFT alone faces practical obstacles for multi-page analysis, including tight token budgets, unstable cross-page references, tiling artifacts, and rising computation~\cite{xing-etal-2025-intelligent}. These limitations motivate a unified approach that aligns visual--textual representation, reconstructs cross-page dependencies into a global document tree, and exploits that structure for chunking. In response, we propose \textbf{\ours}, which integrates SharedDet-based DP/OCR, LVLM-based Multi-modal block embeddings, biaffine dependency scoring~\cite{dozat2017deep}, and tree-guided chunk assembly.

\begin{figure*}[ht!]
\begin{center}
\includegraphics[width=0.85\linewidth]{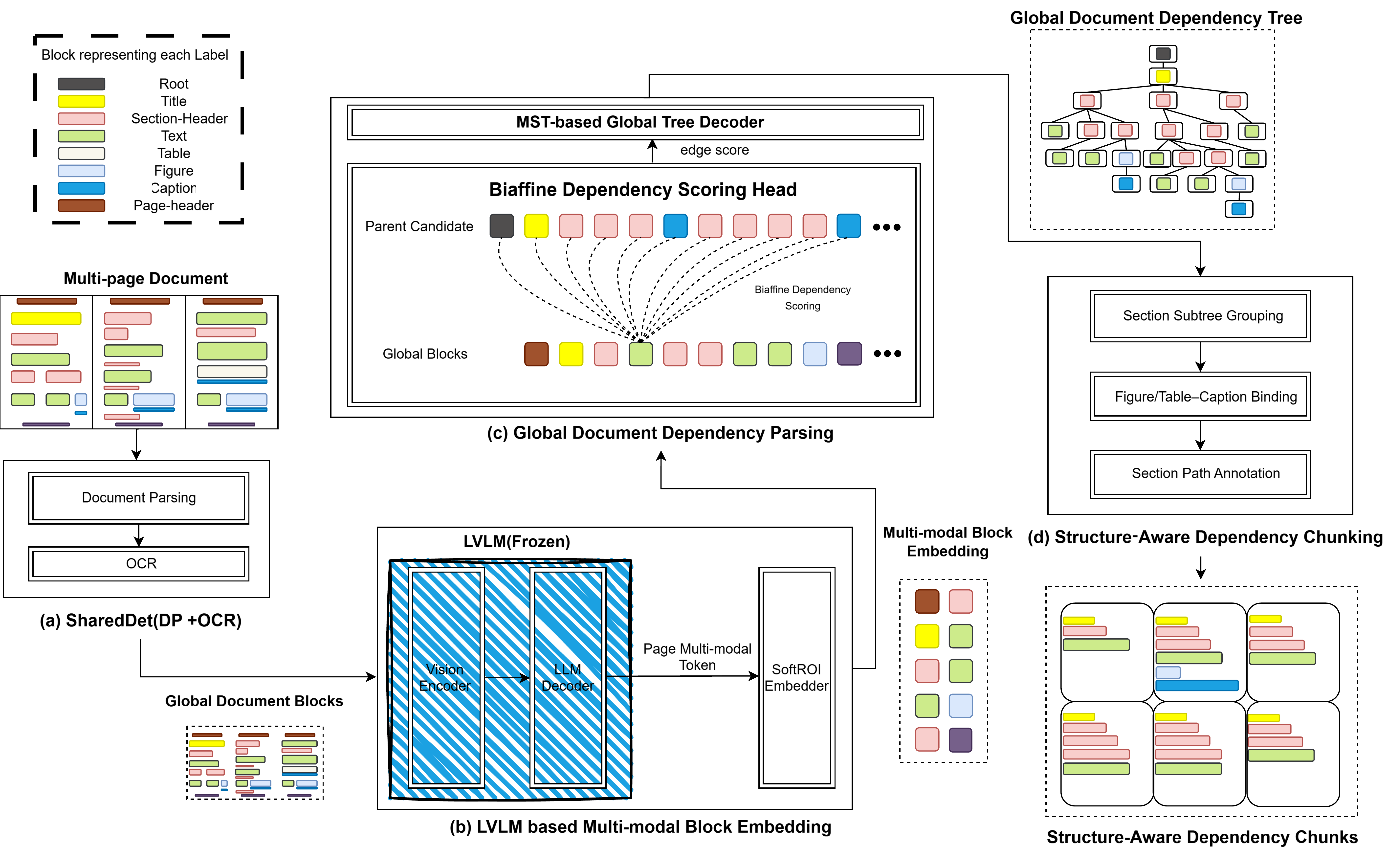}
\end{center}
\caption{Overview of \textbf{\ours}. (a) SharedDet (DP+OCR) converts multi-page documents into Global Document Blocks $\mathcal{V}$. (b) A frozen LVLM with SoftROI pooling produces multi-modal block embeddings $e_i$. (c) A biaffine scorer and MST decoder recover a global document dependency tree $\mathcal{T}$. (d) Structure-Aware Dependency Chunking deterministically converts $\mathcal{T}$ into chunks $\mathcal{C}$ with section paths and page spans.}

\label{fig:system_architecture}

\end{figure*}

\section{\textbf{\ours}}
\label{sec:method}
\textbf{\ours} (Multi-modal, Multi-page, Multi-document Dependency Chunking with Large Vision--Language Models) is a parse-then-chunk pipeline for long industrial documents. Its central idea is to recover block dependencies before constructing retrieval units, so that chunk boundaries follow document structure rather than surface text alone.

As illustrated in Fig.~\ref{fig:system_architecture}, the pipeline consists of four stages: (a) SharedDet (DP+OCR), which converts pages into a shared block canvas $\mathcal{V}$; (b) LVLM-based Multi-modal Block Embedding, which maps each block to a multi-modal embedding $e_i$; (c) Global Document Dependency Parsing, which scores candidate parent--child edges and decodes a global tree $\mathcal{T}$; and (d) Structure-Aware Dependency Chunking, which deterministically converts $\mathcal{T}$ into chunks $\mathcal{C}$ annotated with section paths and page spans. In this work, ``multi-document'' refers to a corpus-level retrieval setting in which dependency trees are constructed per document and chunk indices are queried jointly across documents.

\paragraph{Notation}
Across stages~(a)--(d), we denote the page set by \(D=\{P_t\}_{t=1}^{T}\), the SharedDet blocks by \(\mathcal{V}=\{v_i\}_{i=1}^{N}\) with \(v_i=(\widehat{\mathrm{bbox}}_i,\widehat{\mathrm{type}}_i,\mathrm{text}_i,t(i))\), the block and type embeddings by \(e_i\) and \(\tau_i\), the candidate parent set of child \(v\) by \(\mathcal{P}(v)\), the virtual root by \(r\), the decoded dependency tree by \(\mathcal{T}\), and the final chunk set by \(\mathcal{C}=\{c_m\}_{m=1}^{M}\).

\subsection{(a) SharedDet (DP+OCR)}
\label{sec:stage-a}

\paragraph{Goal}
SharedDet serves as a shared preprocessing protocol rather than the main modeling contribution. We run a fixed DP and OCR pipeline once per document to obtain a stable set of layout blocks, map them into a unified document-level coordinate frame, and reuse these \emph{Global Document Blocks} across all stages of \ours and all compared chunking methods.

\paragraph{Inputs/Outputs}
Given page images \(D=\{P_t\}_{t=1}^{T}\), SharedDet outputs global blocks \(\mathcal{V}=\{v_i\}_{i=1}^{N}\). Each block \(v_i=(\widehat{\mathrm{bbox}}_i,\widehat{\mathrm{type}}_i,\mathrm{text}_i,t(i))\) stores a globally normalized bounding box \(\widehat{\mathrm{bbox}}_i\in[0,1]^2\!\times\![0,1]^2\), a normalized block type \(\widehat{\mathrm{type}}_i\), OCR text \(\mathrm{text}_i\), and page index \(t(i)\in\{1,\dots,T\}\). Bounding boxes and page IDs allow downstream stages to recover the corresponding image and table regions.

\paragraph{Process}
\begin{enumerate}
\item \textbf{Layout detection.} A frozen detector (e.g., DETR/DiT/VGT) extracts block candidates on each page, refined by a confidence threshold \(\tau_{\mathrm{det}}\), non-maximum suppression (NMS) with IoU threshold \(\tau_{\mathrm{nms}}\), and a per-page upper bound \(K_{\max}\). Freezing DP stabilizes block anchors and prevents detector updates from changing the shared input.

\item \textbf{Block OCR.} Each detected block is processed by a dedicated OCR engine (e.g., Tesseract/EasyOCR/TrOCR) to obtain \(\mathrm{text}_i\).
\end{enumerate}

\paragraph{Advantages}
SharedDet provides (1) backbone-invariant reproducibility through a fixed DP+OCR output, (2) global geometric consistency via a document-level coordinate frame, (3) robustness to OCR noise by operating at the block level, (4) high scalability through page-level parallelism, and (5) preservation of image/table regions for downstream Multi-modal chunking.

\subsection{(b) LVLM-based Multi-modal Block Embedding}
\label{sec:stage-b}

\paragraph{Goal}
Given the Global Document Blocks from stage~(a), this stage uses a \emph{frozen} LVLM backbone to convert each page into \emph{Page Multi-modal Tokens}, and then applies a \emph{SoftROI Embedder} to obtain multi-modal block embeddings.

\paragraph{Inputs/Outputs}
The inputs are (i) page images and (ii) the Global Document Blocks \(\mathcal{V}\) from stage~(a).
First, the LVLM produces \emph{Page Multi-modal Tokens} for each page.
Then, the SoftROI Embedder takes these tokens together with the block boxes \(\widehat{\mathrm{bbox}}_i\) and outputs, for each block \(i\), a \emph{SoftROI multi-modal Block Embedding} \(e_i\).
The normalized block type is further embedded via a small lookup table to obtain a type embedding \(\tau_i\), which is used jointly with \(e_i\) in the subsequent dependency scoring head.

\paragraph{Process}
\begin{enumerate}
\item \textbf{Page Multi-modal Tokens from a LVLM.}
Each page image is fed into a \emph{frozen} LVLM (e.g., Qwen2.5-VL, LLaVA-OneVision) with a vision encoder and LLM decoder. From the final decoder layer, we extract the hidden states at positions corresponding to image tokens (Multi-modal representation); these are the \emph{Page Multi-modal Tokens} shown in Fig.~\ref{fig:system_architecture}(b).
Using token-grid metadata, we assign each token \([0,1]^2\) coordinates and apply the global mapping from \S\ref{sec:stage-a} to place them in a unified document-level coordinate frame.

\item \textbf{SoftROI Embedder: SoftROI multi-modal Block Embeddings.}
The \emph{SoftROI Embedder} consumes the Page Multi-modal Tokens and the Global Document Blocks.
For each block \(i\) with global box \(\widehat{\mathrm{bbox}}_i\), we collect all LVLM tokens \(p \in \mathrm{ROI}_i\) whose document-level coordinates fall inside \(\widehat{\mathrm{bbox}}_i\).
Each token receives a boundary-aware weight
{\small
\begin{equation}
\begin{aligned}
w_p \propto \bigl(u_p(1{-}u_p)\bigr)^{\alpha}\bigl(v_p(1{-}v_p)\bigr)^{\alpha},
\qquad \tilde w_p=\frac{w_p}{\sum_{q\in \mathrm{ROI}_i} w_q}
\end{aligned}
\label{eq:eq2}
\end{equation}
}
where \((u_p,v_p)\) are normalized box coordinates and \(\alpha\) is a boundary-sharpening exponent.
The resulting \emph{SoftROI Multi-modal Block Embedding} is
\begin{equation}
\begin{aligned}
e_i=\sum_{p\in \mathrm{ROI}_i}\tilde w_p\,z_p,
\end{aligned}
\label{eq:eq3}
\end{equation}
where \(z_p\) denotes the Page Multi-modal Token at position \(p\).
This boundary-aware pooling adapts RoIAlign-style continuous sampling~\cite{he2017maskrcnn} to the document token grid while retaining the flexibility of deformable RoI pooling~\cite{dai2017deformable}.
To inject layout-aware priors, we also look up a compact type embedding \(\tau_i\) 
from the normalized layout label and combine it with \(e_i\); this type-aware block 
embedding is then fed into the Biaffine Dependency Scoring Head in stage~(c).
\end{enumerate}

\paragraph{Advantages}
Compared to uniform pooling, the boundary-aware weighting is more robust to annotation and border noise; compared to fully deformable RoI pooling, it reduces compute and memory cost while still respecting box geometry~\cite{he2017maskrcnn,dai2017deformable}.
By converting pages into LVLM-derived multi-modal tokens and pooling them with the SoftROI Embedder, this stage produces strong block-level features for downstream dependency scoring.

\subsection{(c) Global Document Dependency Parsing} \label{sec:stage-c} \paragraph{Goal} Score parent--child dependencies across pages and recover a globally valid document tree that satisfies single-root, single-parent, and acyclicity constraints.

\paragraph{Inputs/Outputs} The inputs are multimodal block embeddings \(\{e_i\}\) from stage~(b) and optional type embeddings \(\{\tau_i\}\). The output is a directed dependency tree \(\mathcal{T}\) over all block nodes.

\paragraph{Process}
\begin{enumerate}
\item \textbf{Biaffine Dependency Scoring Head.}
The scorer operates on the global blocks and their SoftROI multi-modal embeddings (Fig.~\ref{fig:system_architecture}(c), ``Global Blocks'').
For each block \(v\in\mathcal{V}\), we construct a small set of \emph{parent candidates} \(\mathcal{P}(v)\) by prioritizing title and section headers, allowing upward links within a column with a small \(y\)-tolerance, and restricting cross-page parents to the most recent \(M\) pages.
We keep the top \(k\) candidates for each child \(v\) according to vertical distance and a header prior, as illustrated by the ``Parent Candidate'' row in Fig.~\ref{fig:system_architecture}(c).
Thus, each child chooses from a small plausible set \(\mathcal{P}(v)\cup\{r\}\): compatible headers, captions, or the virtual root \(r\), which starts a top-level subtree. For each node \(i\), we form an input vector
\(
x_i = [\,e_i;\tau_i\,]
\)
and obtain a hidden representation \(h_i\) via a small MLP.
The Biaffine Dependency Scoring Head then assigns a score to a candidate edge \(u\!\to\!v\) as
\begin{equation}
\begin{aligned}
s(u\!\to\! v)
= [h_u;1]^{\!\top}U[h_v;1]
+ w_{\mathrm{geo}}^{\top}\delta g(u,v),
\end{aligned}
\label{eq:eq4}
\end{equation}
where \(\delta g(u,v)\) encodes pairwise geometric features (normalized relative offsets, block-size ratios, page distance, overlap indicators, and so on).
The score for the virtual root is defined as
\(
s(\text{ROOT}\!\to\! v)= r^{\top} h_v + b_r.
\)

\item \textbf{MST-based Global Tree Decoder.}
Given the edge scores \(s(u\!\to\!v)\) produced by the Biaffine Dependency Scoring Head,
for each child \(v\) we first normalize the scores over its parent candidates
and the virtual root with a \(K{+}1\) child-softmax,
\begin{equation}
\begin{aligned}
P_\theta(p\,|\,v)
= \frac{\exp s(p\!\to\! v)}{\sum_{q\in \mathcal{P}(v)\cup\{r\}} \exp s(q\!\to\! v)},
\end{aligned}
\label{eq:eq5}
\end{equation}
and train the model by minimizing cross-entropy against the ground-truth parent
\(p^{\star}(v)\).
At inference time, independently taking the best-scoring parent for each child can create cycles or mutually inconsistent links. We therefore treat the edge scores \(s(p\!\to\! v)\) (including ROOT edges) as weights and pass them to the \emph{MST-based Global Tree Decoder}~\cite{chuliu1965,edmonds1967}, which returns the highest-scoring globally compatible tree \(\mathcal{T}\).
For comparison, we also report a local argmax baseline that independently selects the best-scoring parent for each child without enforcing global tree constraints.
\end{enumerate}

\paragraph{Advantages}
This stage is the core of \ours: it replaces long-form sequence generation with \emph{direct edge-wise dependency scoring} and graph decoding, so that document structure is modeled explicitly as a scored tree over blocks rather than as a fragile token sequence.
Header-centric, type-aware candidate filtering encodes strong structural priors that rule out implausible parents and focus learning on realistic attachments among headers, captions, and ROOT.
On top of these filtered edges, a biaffine scorer over multi-modal block embeddings, trained with a \(K{+}1\) child-softmax, calibrates scores over ``candidate parents + ROOT'' for each block.
Finally, the MST-based Global Tree Decoder turns these scores into a single, globally consistent document dependency tree that satisfies single-root, single-parent, and acyclicity constraints, yielding interpretable hierarchies that SFT-based LVLMs struggle to recover. Because the decoded tree is guaranteed to be single-rooted and acyclic, downstream chunk construction operates on a logically consistent structure rather than on independently selected local links.

\subsection{(d) Structure-Aware Dependency Chunking}
\label{sec:stage-d}

\paragraph{Goal}
Map the decoded dependency tree into retrieval units that preserve section/subsection boundaries and keep figures or tables attached to their captions, so that each chunk remains structurally coherent even across page breaks.

\paragraph{Inputs/Outputs}
Given the decoded dependency tree \(\mathcal{T}\) over block nodes, chunking deterministically produces \emph{Structure-Aware Dependency Chunks} \(\mathcal{C}=\{c_m\}_{m=1}^{M}\). Each chunk \(c_m=(B_m,\pi_m,[p_m^{\min},p_m^{\max}])\) contains a subset of blocks \(B_m\), its root-to-section path \(\pi_m\), and the covered page span; block IDs and layout metadata are retained for retrieval-time serialization.

\paragraph{Process}
\begin{enumerate}
\item \textbf{Section-root DFS.}
We traverse \(\mathcal{T}\) from title and section-header nodes, treating each such node as a structural anchor.
A DFS collects its descendants into a section subtree, merging blocks across page boundaries whenever they belong to the same subtree.

\item \textbf{Visual-text binding.}
If a figure/table node and a caption node are linked in \(\mathcal{T}\), we force them into the same block subset \(B_m\); if the edge is missing, we fall back to the closest compatible pair by spatial proximity.

\item \textbf{Chunk emission.}
For each retained subtree or merged visual-text group, we emit a chunk together with its section path, page span, and constituent block list.
This makes chunk construction a deterministic post-processing step on \(\mathcal{T}\), rather than a separate learned module.
\end{enumerate}

\paragraph{Advantages}
Because chunk boundaries follow \(\mathcal{T}\), section continuations are not arbitrarily split and cross-page evidence stays connected. Visual-text binding keeps figures and tables with their descriptions, while the emitted section path and page span make otherwise similar chunks distinguishable in a \emph{multi-document} index. The result is a retrieval unit that aligns more closely with the document's true semantic and hierarchical structure, improving both retrieval precision and downstream QA. Chunk granularity is still controlled deterministically on the recovered tree by adjusting the maximum chunk length and cut policies, enabling section-level, paragraph-level, or finer chunking without retraining the parser.

\begin{table*}[t!]
\centering
\footnotesize
\renewcommand{\arraystretch}{1.1}
\setlength{\tabcolsep}{4pt}
\resizebox{0.65\textwidth}{!}{%
\begin{tabular}{l cc cc cc}
\toprule
\multirow{2}{*}{\textbf{Method (Setting)}} &
\multicolumn{2}{c}{\textbf{HRDS}} &
\multicolumn{2}{c}{\textbf{HRDH}} &
\multicolumn{2}{c}{\textbf{DocHieNet}} \\
\cmidrule(lr){2-3}\cmidrule(lr){4-5}\cmidrule(lr){6-7}
& \textbf{F1} & \textbf{STEDS}
& \textbf{F1} & \textbf{STEDS}
& \textbf{F1} & \textbf{STEDS} \\
\midrule
\multicolumn{7}{l}{\emph{Image understanding LVLM}}\\[0.15em]
GPT-5 & 35.39 & 26.27 & 32.03 & 24.52 & 29.12 & 21.94 \\
LLaVA-OneVision-1.5 & 27.61 & 12.93 & 26.30 & 18.21 & 17.78 & 8.57 \\
InternVL-3.5       & 28.40 & 14.47 & 27.57 & 19.98 & 18.18 & 9.60 \\
Qwen2.5-VL          & 28.41 & 14.51 & 27.57 & 19.99 & 18.20 & 9.62 \\
\midrule
\multicolumn{7}{l}{\emph{Tree-aware models \;\; (shared GT layout)}}\\[0.15em]
DocParser                     & 47.09 & 31.03 & 35.41 & 27.15 & 10.68 & 4.31 \\
DSG                           & 48.43 & 32.13 & 36.42 & 27.69 & 26.71 & 19.45 \\
DSPS                          & 65.27 & 59.57 & 54.06 & 38.41 & 35.61 & 23.81 \\
DSHP-LLM                      & 44.90 & 29.52 & 61.29 & 51.34 & 64.29 & 53.49 \\
Qwen2.5-VL\textendash DHP\textendash SFT & 50.97 & 46.75 & 43.05 & 41.02 & 42.85 & 40.39 \\
\rowcolor{blue!6}
\textbf{\ours}  & \textbf{82.87} & \textbf{76.52} & \textbf{77.75} & \textbf{71.65} & \textbf{76.01} & \textbf{70.83} \\
\bottomrule
\end{tabular}%
}
\caption{Performance (\%) on DHP datasets (HRDS, HRDH, DocHieNet). Tree-aware baselines and M3DocDep are evaluated on the same GT layout blocks, so that results isolate hierarchy recovery rather than document parsing quality; general-purpose LVLM rows are included as reference baselines. Implementation details are given in the Supplementary. Boldface numbers indicate the best performance in each column.}
\label{tab:hierarchy_results}
\end{table*}

\section{Experimental Settings}
This section summarizes the training setup for \textbf{\ours} on hierarchical and dependency structure, as well as the evaluation protocol for RAG-based VQA. Additional dataset statistics, hyperparameters, and hardware details are provided in the Supplementary.

\paragraph{Datasets}
For hierarchical/dependency parsing, we train and evaluate on DocHieNet~\cite{xing2024dochienet} and HRDH/HRDS~\cite{10.1609/aaai.v37i2.25277}, which cover diverse document domains and layouts.
For RAG-based VQA, we evaluate on DUDE~\cite{10376937}, MP-DocVQA~\cite{tito2023hierarchicalmultimodaltransformersmultipage}, CUAD~\cite{hendrycks2021cuad}, and MOAMOB~\cite{hong-etal-2024-intelligent}, spanning financial reports, contracts, scanned forms, and complex structured documents. All test documents are jointly indexed at the corpus level, and the top \(k_{\mathrm{ret}}\) retrieved chunks (default
\(k_{\mathrm{ret}}{=}4\)) are used to generate answers. This corpus-level retrieval reflects real-world deployment and increases cross-document disambiguation difficulty.

\paragraph{Shared evaluation protocol}
Unless otherwise stated, all chunking methods operate on the same Global Document Blocks produced by SharedDet and use the same maximum chunk length (550 tokens). Retrieval is performed at the corpus level over all test documents, with \(k_{\mathrm{ret}} \in \{1,2,3,4\}\) and default \(k_{\mathrm{ret}}{=}4\). Each chunk is serialized under a shared schema consisting of section path, page range, block-type markers, and OCR/caption text; fields unavailable to a given chunker are left blank or omitted. Figure/table chunks keep the associated caption in the same retrieval unit. Text-only retrievers (BGE, E5, BM25) operate on this shared serialized text, while MM-Embed additionally receives the associated figure/table crops when present. For LVLM readers, retrieved chunks are provided under the same schema, with figure/table crops passed alongside the serialized text when available. This protocol is intended to isolate chunk quality rather than differences in parsing, chunk budget, retriever, or reader configuration. We additionally report a pairwise no-metadata fairness control for MultiDocFusion and \ours that removes section-path and page-range fields from both methods, isolating boundary quality from any gain due purely to metadata.

\begin{table*}[t!]
\centering
\small
\renewcommand{\arraystretch}{1.25}
\setlength{\tabcolsep}{3.5pt}
\resizebox{\textwidth}{!}{%
\begin{tabular}{l ccc ccc ccc ccc}
\toprule
\multirow{2}{*}{\textbf{Chunking Method}}
  & \multicolumn{3}{c}{\textbf{DUDE}}
  & \multicolumn{3}{c}{\textbf{MP-DocVQA}}
  & \multicolumn{3}{c}{\textbf{CUAD}}
  & \multicolumn{3}{c}{\textbf{MOAMOB}} \\
\cmidrule(lr){2-4} \cmidrule(lr){5-7} \cmidrule(lr){8-10} \cmidrule(lr){11-13}
 & R & P & nDCG
 & R & P & nDCG
 & R & P & nDCG
 & R & P & nDCG \\
\midrule
Length chunking
  & 26.28 & 16.86 & 21.66
  & 25.23 & 15.87 & 19.33
  & 90.11 & 85.37 & 87.76
  & 64.62 & 56.76 & 62.09 \\
Semantic chunking
  & 9.56 & 5.49 & 7.75
  & 9.39 & 5.24 & 6.80
  & 76.84 & 67.19 & 71.81
  & 27.37 & 19.50 & 24.53 \\
LumberChunker
  & 23.95 & 15.33 & 19.86
  & 21.52 & 12.98 & 16.09
  & 90.31 & 85.76 & 88.00
  & 61.30 & 52.05 & 56.92 \\
Perplexity chunking
  & 24.28 & 15.59 & 20.20
  & 21.59 & 13.18 & 16.29
  & 88.69 & 83.95 & 86.03
  & 61.73 & 52.41 & 57.85 \\
Structure-based chunking
  & 22.19 & 14.50 & 18.62
  & 20.36 & 12.30 & 15.24
  & 88.44 & 83.11 & 85.81
  & 55.44 & 46.62 & 51.49 \\
MultiDocFusion
  & 29.27 & 20.01 & 25.05
  & 27.05 & 17.59 & 21.31
  & 90.21 & 86.51 & 88.19
  & 67.58 & 61.84 & 65.54 \\
\rowcolor{blue!6}
\textbf{\ours}
  & \textbf{35.12} & \textbf{24.91} & \textbf{27.81}
  & \textbf{31.28} & \textbf{19.76} & \textbf{24.52}
  & \textbf{91.25} & \textbf{87.23} & \textbf{89.12}
  & \textbf{76.97} & \textbf{72.83} & \textbf{75.54} \\
\bottomrule
\end{tabular}%
}
\caption{Retrieval performance (\%) by chunking method (macro-averaged Recall, Precision, and nDCG for top-$k \in \{1,2,3,4\}$) on DUDE, MP-DocVQA, CUAD, and MOAMOB. Results are averaged over BGE, E5, BM25, and MM-Embed retrievers under the shared corpus-level RAG protocol. Best scores are in \textbf{bold}.}
\label{tab:chunking_results_four_datasets}
\end{table*}

\paragraph{Models}
\textbf{\ours} attaches a biaffine relation head to multi-modal LVLM backbones---LLaVA-OneVision-1.5~\cite{an2025llavaonevision15fullyopenframework}, InternVL-3.5~\cite{wang2025internvl35advancingopensourcemultimodal}, and Qwen2.5-VL~\cite{bai2025qwen25vltechnicalreport}---to reconstruct block-level hierarchies. We compare against a Qwen2.5-VL--DHP--SFT baseline and prior DHP baselines (DocParser~\cite{rausch2021docparser}, DSG~\cite{rausch2023dsgendtoenddocumentstructure}, DSPS~\cite{10.1609/aaai.v37i2.25277}, DSHP-LLM~\cite{multidocfusion}), as well as Length~\cite{Gong2020Recurrent}, Semantic~\cite{qu-etal-2025-semantic}, LumberChunker~\cite{duarte2024lumberchunker}, Perplexity~\cite{zhao2024metachunking}, Structure-based~\cite{yepes2024financialreportchunkingeffective}, and MultiDocFusion~\cite{multidocfusion} for chunking. Additional backbone swaps are reported in the Supplementary.

\paragraph{Evaluation}
We report F1 (parent prediction) and STEDS (tree reconstruction)~\cite{DBLP:journals/corr/abs-1805-06869} for hierarchy/dependency recovery. Retrieval quality is measured with Precision, Recall, and nDCG~\cite{jarvelin2002cumulated}, and VQA answers are evaluated using ANLS~\cite{biten2019scene}, ROUGE-L~\cite{lin2004rouge}, and METEOR~\cite{banerjee2005meteor}. Table~\ref{tab:hierarchy_results} isolates dependency recovery by evaluating all tree-aware methods on the same GT layout blocks, while the retrieval and QA tables evaluate chunk quality under the shared corpus-level RAG protocol.

\begin{table*}[t!]
\centering
\small
\renewcommand{\arraystretch}{1.25}
\setlength{\tabcolsep}{3.5pt}
\resizebox{\textwidth}{!}{%
\begin{tabular}{l ccc ccc ccc ccc}
\toprule
\multirow{2}{*}{\textbf{Chunking Method}}
 & \multicolumn{3}{c}{\textbf{DUDE}}
 & \multicolumn{3}{c}{\textbf{MP-DocVQA}}
 & \multicolumn{3}{c}{\textbf{CUAD}}
 & \multicolumn{3}{c}{\textbf{MOAMOB}} \\
\cmidrule(lr){2-4}\cmidrule(lr){5-7}\cmidrule(lr){8-10}\cmidrule(lr){11-13}
 & ANLS & R-L & MTR & ANLS & R-L & MTR & ANLS & R-L & MTR & ANLS & R-L & MTR \\
\midrule
Length chunking
 & 16.11 & 14.44 & 19.88
 & 13.98 & 9.66 & 14.08
 & 25.85 & 16.77 & 16.62
 & 24.97 & 8.23 & 11.15 \\
Semantic chunking
 & 15.48 & 12.61 & 16.57
 & 13.32 & 8.05 & 9.78
 & 25.93 & 14.91 & 14.68
 & 24.55 & 8.46 & 10.43 \\
LumberChunker
 & 15.31 & 12.84 & 17.52
 & 13.07 & 7.69 & 9.93
 & 26.57 & 16.30 & 16.50
 & 25.36 & 8.48 & 11.67 \\
Perplexity chunking
 & 16.53 & 13.90 & 18.55
 & 13.44 & 7.51 & 9.50
 & 26.41 & 16.46 & 15.24
 & 25.32 & 8.94 & 11.90 \\
Structure-based chunking
 & 17.51 & 14.89 & 19.21
 & 15.37 & 9.80 & 12.78
 & 24.98 & 15.56 & 15.91
 & 25.01 & 9.79 & 11.14 \\
MultiDocFusion
 & 18.59 & 16.92 & 22.85
 & 16.15 & 13.16 & 18.50
 & 27.38 & 17.62 & 16.50
 & 25.96 & 9.16 & 12.57 \\
\rowcolor{blue!6}
\textbf{\ours}
 & \textbf{21.43} & \textbf{18.21} & \textbf{25.12}
 & \textbf{18.17} & \textbf{15.29} & \textbf{20.14}
 & \textbf{29.25} & \textbf{19.78} & \textbf{18.23}
 & \textbf{27.14} & \textbf{10.22} & \textbf{14.68} \\
\bottomrule
\end{tabular}%
}
\caption{Average QA performance (\%) of chunking strategies on DUDE, MP-DocVQA, CUAD, and MOAMOB for top-$k \in \{1,2,3,4\}$. R-L = ROUGE-L, MTR = METEOR. Results are averaged over LLaVA-OneVision-1.5, InternVL-3.5, and Qwen2.5-VL readers under the shared corpus-level RAG protocol.}
\label{tab:chunking_vs_datasets}
\end{table*}

\section{Experimental Results}
\label{sec:experimental_results}

We evaluate \textbf{\ours} on three axes: hierarchy recovery, retrieval quality, and downstream QA. Robustness to alternative DP/OCR/retrieval modules, additional LVLM swaps, and runtime analyses are deferred to the Supplementary.

\subsection{DHP Performance in Different Methods}
\label{subsec:DHP_performance}
Table~\ref{tab:hierarchy_results} shows that \textbf{\ours} outperforms both prior tree-aware baselines and general-purpose LVLM reference baselines on HRDS, HRDH, and DocHieNet when all methods operate on the same GT layout blocks, indicating better local parent prediction and better global tree consistency. The gains are especially large on HRDH and DocHieNet, where diverse layouts and cross-page relations make hierarchy recovery harder. Across alternative LVLM encoders for block embeddings, M3DocDep remains stable on DocHieNet parent prediction, achieving 76.01 / 75.71 / 74.07 F1 with Qwen2.5-VL, InternVL-3.5, and LLaVA-OneVision-1.5, respectively (Supplementary).

\subsection{Performance in Different Chunking Methods}
Table~\ref{tab:chunking_results_four_datasets} reports macro-averaged retrieval performance across four multi-page VQA corpora. \textbf{\ours} achieves the best results on all datasets and metrics, with the largest gains on DUDE, MP-DocVQA, and MOAMOB, where cross-page evidence, OCR noise, and figure/table regions make retrieval especially sensitive to chunk boundaries. This supports the intended causal chain: better dependency recovery yields cleaner chunk boundaries, which in turn improves both evidence coverage and ranking quality. Additional qualitative examples, stronger backbone swaps, and failure cases are provided in the Supplementary.

Figure~\ref{fig:qualitative_e2e} traces a representative document through the \textbf{\ours} pipeline: from a 5-page input, through the recovered dependency subtree, to a structure-aware chunk that binds a figure with its caption under the correct section path.

\begin{figure*}[t]
\centering

% --- (a) page thumbnails ---
\begin{minipage}[b]{0.12\textwidth}
  \centering
  \includegraphics[width=\linewidth,height=0.30\textheight,keepaspectratio]{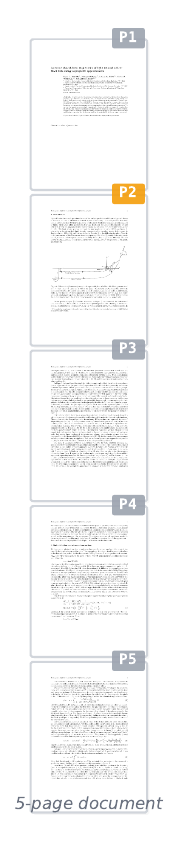}\\[4pt]
  \makebox[0pt]{\small (a) Multi-page input\,}
\end{minipage}%
\hfill
{\large$\;\boldsymbol{\rightarrow}\;$}%
\hfill
% --- (b) cropped dependency tree ---
\begin{minipage}[b]{0.52\textwidth}
  \centering
  \includegraphics[width=\linewidth]{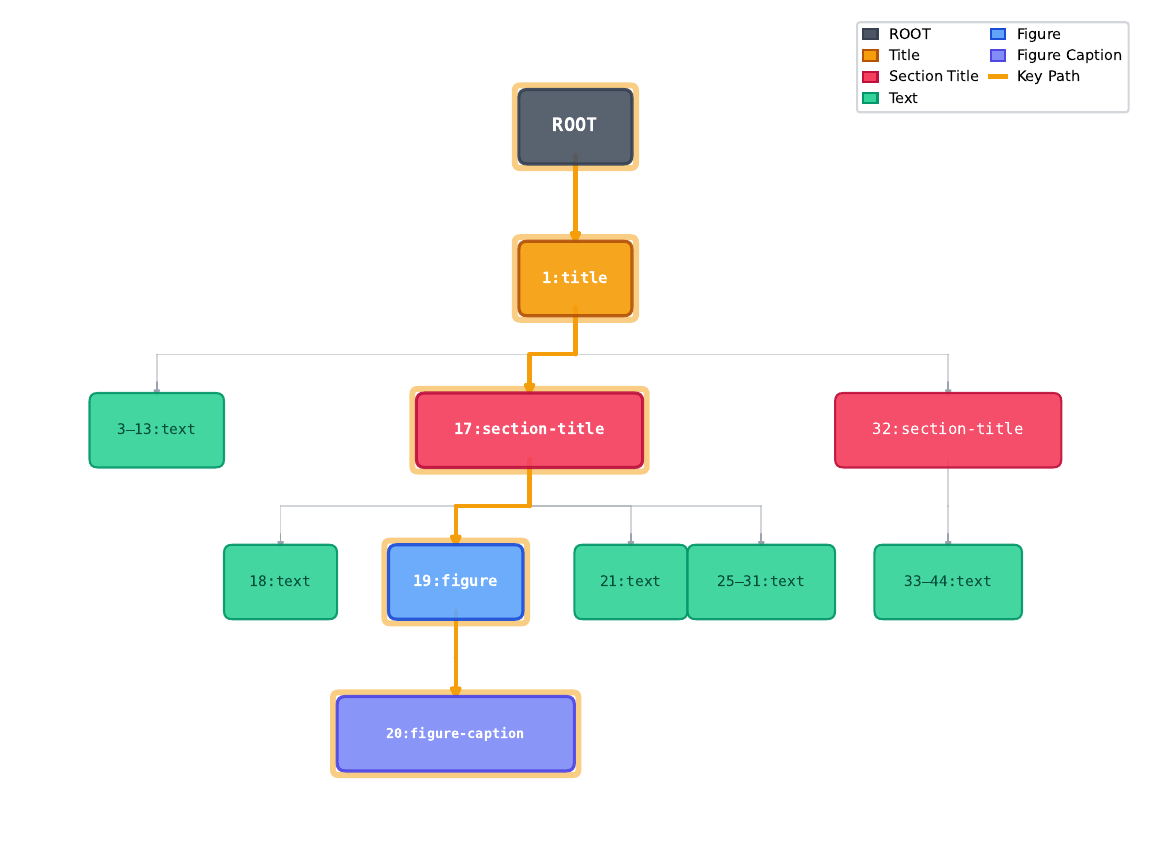}\\[4pt]
  {\small (b) Recovered dependency subtree from $\mathcal{T}$}
\end{minipage}%
\hfill
{\large$\boldsymbol{\rightarrow}$}%
\hfill
% --- (c) chunk card ---
\begin{minipage}[b]{0.27\textwidth}
  \centering
  \includegraphics[width=\linewidth,height=0.30\textheight,keepaspectratio]{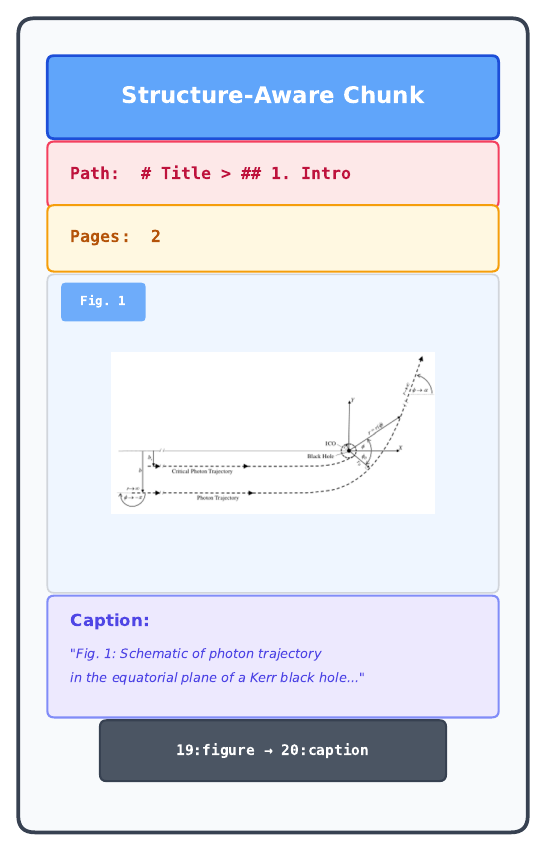}\\[4pt]
  {\small (c) Output chunk $c_m$}
\end{minipage}

\caption{End-to-end qualitative example of \textbf{\ours}.
(a)~A 5-page industrial document is input.
(b)~The recovered dependency subtree (cropped from full tree~$\mathcal{T}$): \texttt{1:title}~$\to$~\texttt{17:section-title}~$\to$~\texttt{19:figure}~$\to$~\texttt{20:figure-caption} shows the figure--caption binding under the governing section.
(c)~Structure-aware chunking emits a chunk that keeps the figure crop and its caption together, annotated with the section path and page span.
The full tree and additional examples are in the Supplementary.}
\label{fig:qualitative_e2e}
\end{figure*}

\subsection{QA Performance in Chunking Methods}
\label{subsec:qa_performance}
Table~\ref{tab:chunking_vs_datasets} shows that the retrieval gains translate into downstream QA improvements across all four corpora. \textbf{\ours} consistently improves ANLS and also improves overlap-based metrics, indicating that tree-guided chunking helps package evidence in a way that better matches document structure. In a pairwise no-metadata fairness control that removes section-path and page-range fields from both MultiDocFusion and \ours, our method still retains a 2.3\% nDCG advantage, indicating that the gains are driven primarily by better chunk boundaries rather than by metadata alone.

\subsection{Ablation Studies}

\begin{table}[t]
\centering
\small
\renewcommand{\arraystretch}{1.25}
\setlength{\tabcolsep}{5pt}
\begin{tabular}{lcc}
\toprule
\textbf{Variant} & \textbf{Avg F1} & \textbf{Avg STEDS} \\
\midrule
\textbf{Full} & \textbf{78.88} & \textbf{73.00} \\
\midrule
MST $\rightarrow$ local argmax & 73.68 \textcolor{red}{\small($-$5.19)} & 66.30 \textcolor{red}{\small($-$6.70)} \\
Disallow cross-page edges & 71.73 \textcolor{red}{\small($-$7.15)} & 63.74 \textcolor{red}{\small($-$9.26)} \\
\bottomrule
\end{tabular}
\caption{Condensed ablation (\%) on the two most important design choices, macro-averaged over HRDS, HRDH, and DocHieNet. Full per-dataset ablations are in the Supplementary.}
\label{tab:ablation_condensed}
\end{table}

We evaluate the contribution of each core module in \textbf{\ours} under the SharedDet setting. As shown in Table~\ref{tab:ablation_condensed}, the largest degradations occur when cross-page edges are disabled or when MST decoding is replaced with local \texttt{argmax}, reducing macro-averaged F1/STEDS by 7.15/9.26 and 5.19/6.70 points, respectively. This shows that preserving cross-page links and enforcing global tree constraints is crucial for stable hierarchy recovery. Full per-dataset ablations, including SoftROI, header-centric priors, and candidate top-$k$ pruning, are deferred to the Supplementary.

\section{Conclusion}
\label{sec:conclusion}

\textbf{\ours} improves multi-page, multi-document RAG by explicitly recovering block dependencies before chunk construction. Instead of relying on long-form sequence generation to infer structure, it scores candidate parent--child edges over LVLM-based multi-modal block representations, decodes a globally valid tree, and builds tree-guided chunks that preserve section structure and figure/table--caption relations. Across hierarchy recovery, retrieval, and QA, the same pattern recurs: better dependency recovery leads to cleaner chunk boundaries, more coherent retrieval units, and stronger downstream QA. Supplementary results further show that this pattern remains robust under stronger DP/OCR backbones and alternative LVLM encoders. Future work includes joint training of the LVLM and dependency head, lower-cost supervision for tree induction at scale, and lighter-weight variants for latency-sensitive industrial deployment.

\section*{Acknowledgements}
This work was supported by the Commercialization Promotion Agency for R\&D Outcomes (COMPA) grant funded by the Korea government (Ministry of Science and ICT) (2710086166).
This work was supported by Institute for Information \& Communications Technology Promotion (IITP) grant funded by the Korea government (MSIT) (RS-2024-00398115, Research on the reliability and coherence of outcomes produced by Generative AI).
This research was supported by Basic Science Research Program through the National Research Foundation of Korea (NRF) funded by the Ministry of Education (NRF-2022R1A2C1007616).
This work was supported by ICT Creative Consilience Program through the Institute of Information \& Communications Technology Planning \& Evaluation (IITP) grant funded by the Korea government (MSIT) (IITP-2026-RS-2020-II201819).

\bibliographystyle{ieee_fullname}
\bibliography{main}

\clearpage
\appendix
\setcounter{figure}{0}
\setcounter{table}{0}
\setcounter{algorithm}{0}
\renewcommand{\thefigure}{A\arabic{figure}}
\renewcommand{\thetable}{A\arabic{table}}
\renewcommand{\thealgorithm}{A\arabic{algorithm}}
% This appendix complements the main text and provides a
% roadmap for reproducibility and inspection of results. 

% ============================================================
\section{Datasets and Pre-processing Details}
\label{sec:datasets_supp}

All datasets used in our experiments are publicly available research benchmarks. We rely exclusively on open corpora for both hierarchy parsing and RAG-based VQA, and do not use any proprietary or private documents.

\begin{table}[h!]
\centering
\resizebox{\linewidth}{!}{%
\begin{tabular}{lcccc}
\toprule
\textbf{Dataset} & \textbf{Type/Domain} & \textbf{\#Documents} & \textbf{Avg. Pages} & \textbf{\#QA pairs} \\
\midrule
\textbf{DocHieNet} & Mixed (Reports/Papers/Industrial) & 1,673 & 5.3 & -- \\
\textbf{HRDH} & Academic Papers (arXiv, diverse layouts) & 1,500 & 7.1 & -- \\
\textbf{HRDS} & Academic Papers (ACL Anthology, single template) & 1,000 & 10.4 & -- \\
\midrule
\textbf{MP-DocVQA} & General Documents (Multi-page) & 17,000 & 3.4 & 48,000+ \\
\textbf{CUAD} & Legal Documents (Contracts) & 510 & 6.2 & 13,000+ \\
\textbf{DUDE} & Mixed (Financial Reports/Manuals) & 3,000+ & 4.9 & 7,000+ \\
\textbf{MOAMOB} & Industrial Technical Documents & 2 & 35.5 & 71 \\
\bottomrule
\end{tabular}}
\caption{Summary of key datasets used in our experiments. DocHieNet, HRDH, and HRDS are used for document hierarchy parsing, while MP-DocVQA, CUAD, DUDE, and MOAMOB are used for multi-page QA.}
\label{tab:appendix_dataset_summary}
\end{table}

\subsection{Document Hierarchical Parsing Corpora}

We train and evaluate M3DocDep on three document
hierarchical parsing benchmarks: DocHieNet, HRDH, and
HRDS. These corpora collectively cover a wide range of
industrial document types, including scanned reports,
born-digital PDFs, and documents with complex
multi-column layouts, which match the intended deployment
scenarios for industrial RAG.

\paragraph{DocHieNet.}
DocHieNet~\cite{xing2024dochienet} contains 1{,}673
documents from multiple domains (technical reports,
scientific papers, industrial documents), with many
multi-page scanned PDFs. Each document is annotated
with block-level types (titles, section headers, paragraphs,
tables, figures, captions, etc.) and parent--child relations
between blocks, forming a ground-truth hierarchy tree.
Most documents are English or Chinese.

\paragraph{HRDH and HRDS.}
HRDH and HRDS~\cite{10.1609/aaai.v37i2.25277} are two subsets of HRDoc with different layout characteristics. 
HRDS contains 1,000 ACL Anthology papers with a nearly identical template, offering a clean and homogeneous setting. 
HRDH includes 1,500 arXiv papers with highly diverse layouts across many research domains, making it a harder and more realistic benchmark. 
Both provide block-level parent annotations for hierarchy supervision.

Table~\ref{tab:dataset_stats_unified} summarizes the
document- and page-level coverage for the three DHP
corpus, including the proportion of blocks with parents,
and the fraction of intra-page vs.\ cross-page edges.
These statistics highlight that a large portion of relations
are cross-page, particularly in HRDH/DocHieNet, making
global tree reconstruction crucial for realistic DHP.

\begin{table}[t]
\centering
\resizebox{\linewidth}{!}{%
\begin{tabular}{lccc}
\toprule
\textbf{Dataset} &
\textbf{Documents (multi-page)} &
\textbf{GT page coverage (test)} &
\textbf{Parent--child summary (test)} \\
\midrule
\textbf{DocHieNet} &
161 (100\%) &
\makecell[l]{
\textbf{Pages w/ GT} (avg / med / min / max):\\
\quad 14.205 / 9.000 / 3 / 50\\[2pt]
\textbf{Pages total} (avg / med / min / max):\\
\quad 14.354 / 9.000 / 3 / 50\\[2pt]
\textbf{Coverage} (avg / med / min / max):\\
\quad 0.994 / 1.000 / 0.727 / 1.000} &
\makecell[l]{
\textbf{Valid blocks:} 29{,}401\\
\textbf{Parents:} 24{,}050 (81.80\%)\\
\textbf{Edges:} intra 59.36\%, cross 40.64\%\\
\textbf{ROOT:} 18.20\%} \\
\midrule
\textbf{HRDH} &
500 (100\%) &
\makecell[l]{
\textbf{Pages w/ GT} (avg / med / min / max):\\
\quad 14.080 / 12.000 / 2 / 35\\[2pt]
\textbf{Pages total} (avg / med / min / max):\\
\quad 14.086 / 12.000 / 2 / 35\\[2pt]
\textbf{Coverage} (avg / med / min / max):\\
\quad 0.999 / 1.000 / 0.800 / 1.000} &
\makecell[l]{
\textbf{Valid blocks:} 77{,}954\\
\textbf{Parents:} 56{,}455 (72.42\%)\\
\textbf{Edges:} intra 50.58\%, cross 49.42\%\\
\textbf{ROOT:} 27.58\%} \\
\midrule
\textbf{HRDS} &
100 (100\%) &
\makecell[l]{
\textbf{Pages w/ GT} (avg / med / min / max):\\
\quad 10.380 / 10.500 / 5 / 19\\[2pt]
\textbf{Pages total} (avg / med / min / max):\\
\quad 10.380 / 10.500 / 5 / 19\\[2pt]
\textbf{Coverage} (avg / med / min / max):\\
\quad 1.000 / 1.000 / 1.000 / 1.000} &
\makecell[l]{
\textbf{Valid blocks:} 15{,}584\\
\textbf{Parents:} 11{,}156 (71.59\%)\\
\textbf{Edges:} intra 67.27\%, cross 32.73\%\\
\textbf{ROOT:} 28.41\%} \\
\bottomrule
\end{tabular}}
\caption{Summary of page coverage and parent--child relations in DocHieNet, HRDH, and HRDS, based on the test set.}
\label{tab:dataset_stats_unified}
\end{table}

\subsection{RAG-based VQA Corpora}

For RAG-based evaluation we consider four multi-page VQA
corpus: DUDE, MP-DocVQA, CUAD, and MOAMOB.
They span financial reports, contracts, scanned forms, and
complex industrial documentation.

\paragraph{DUDE.}
DUDE~\cite{10376937} includes more than 3{,}000
documents such as annual reports and technical manuals,
with over 7{,}000 QA pairs. The official test-set answers are
hidden on a server; thus we use the validation split for
retrieval metrics and report ANLS on the official test
server in the main paper.

\paragraph{MP-DocVQA.}
MP-DocVQA~\cite{tito2023hierarchicalmultimodaltransformersmultipage}
contains roughly 17{,}000 multi-page documents with more
than 48{,}000 questions (around 2.8 questions per document).
Documents include various scanned and born-digital
government and industrial reports with heterogeneous
layouts.

\paragraph{CUAD.}
CUAD~\cite{hendrycks2021cuad} consists of 510 legal
contracts with over 13{,}000 QA pairs, targeting specific
clauses and legal concepts. We follow prior work and use
the official test split (about 50 documents, 1{,}200 QA) for
evaluation, indexing all test documents jointly.

\paragraph{MOAMOB.}
MOAMOB~\cite{hong-etal-2024-intelligent} is a
small-scale but challenging dataset with two long industrial
documents in Korean and 71 QA pairs about predictive
maintenance. The questions often require cross-page
reasoning and fine-grained reference to operational
guidelines, making it a stress test for structure-aware
chunking and retrieval.

\subsection{Pre-processing Pipeline}
\label{sec:preprocess_supp}

All documents are processed through the SharedDet
(DP+OCR) pipeline described in Sec.3.1 of the main paper.
Here we detail design choices and hyperparameters.

\paragraph{Document Parsing (DP).}
We use detectors (DETR, DiT, VGT) trained on
DocLayNet~\cite{10.1145/3534678.3539043} to detect
layout blocks (titles, headers, paragraphs, tables, figures,
etc.) per page. For each detector we fix:
(i) a confidence threshold $\tau_{\text{det}}$,
(ii) a NMS IoU threshold $\tau_{\text{nms}}$, and
(iii) a per-page upper bound $K_{\max}$ on block count.
These are tuned once on a held-out validation subset and
reused for all experiments, ensuring that all methods
(M3DocDep and baselines) operate on the same block set.

\paragraph{OCR.}
Each detected block is independently passed to an OCR engine (Tesseract, EasyOCR, or TrOCR) depending on the experiment. We use the default English models, add Korean for MOAMOB, and include Chinese OCR models for DocHieNet documents containing Chinese text. All extracted text is normalized by lowercasing, removing control characters, and collapsing whitespace, and is then associated with each block's bounding box and page index.

\paragraph{Global Document Blocks.}
Bounding boxes are mapped into a global normalized
coordinate frame $[0,1]^2 \times [0,1]^2$ using the original
page sizes and page indices, yielding the Global Document
Blocks
\[
  V = \{(\mathrm{bbox}_i, \mathrm{type}_i, \mathrm{text}_i, t(i))\}_i.
\]
These blocks form a stable, detector-and-OCR-agnostic
canvas for both M3DocDep and all tree-aware/text-based
baselines.

\subsection{Additional Statistics for DHP Corpora}

Table~\ref{tab:dataset_stats_unified} provides a more
detailed view of page coverage and parent--child relations
in DocHieNet, HRDH, and HRDS, which is useful when
analyzing cross-page dependencies and ROOT frequency.

% ============================================================

\section{Metric Definitions and Evaluation Protocol}
\label{sec:metric_supp}

\subsection{Hierarchy Metrics}
For DHP, we report (i) parent prediction F1 and (ii) STEDS~\cite{DBLP:journals/corr/abs-1805-06869}.
Parent F1 is computed over all non-ROOT blocks, treating
each predicted parent as a single-label classification target.
STEDS follows the original definition and measures tree-level
edit distance between predicted and ground-truth hierarchies.

\subsection{Retrieval Metrics}
For each question, we treat a chunk as relevant if it contains
the gold answer span (or its annotated supporting block).
Precision@k, Recall@k, and nDCG@k~\cite{jarvelin2002cumulated} are computed at the
question level and macro-averaged over all questions.

\subsection{QA Metrics}
We compute ANLS~\cite{biten2019scene}, ROUGE-L~\cite{lin2004rouge}, and METEOR~\cite{banerjee2005meteor} on normalized
answers (lowercased, extra whitespace removed, punctuation
stripped). For ANLS we follow the official MP-DocVQA
evaluation, using character-level Levenshtein distance.

% ============================================================

\section{Baseline Implementations}
\label{sec:baselines_supp}

Unless otherwise noted, all methods follow the common RAG setup in Sec.~\ref{sec:lvlm_rag_supp}; only DHP and chunking components differ. All baselines considered in this work are instantiated from publicly available implementations or public APIs; we do not rely on any internal or non-releasable systems. Whenever possible, we use the official code released by the original authors, and otherwise provide faithful re-implementations that follow their published descriptions.

\subsection{Document Hierarchical Parsing (DHP) Baselines}
\label{sec:appendix_dhp_methods}

For DHP baselines (DocParser, DSG, DSPS, DSHP-LLM, Qwen2.5-VL--DHP--SFT),
we either use official implementations or faithful re-implementations following
their papers.

\paragraph{DocParser.}
DocParser~\cite{rausch2021docparser} is a pioneering DHP method
that converts a flat list of layout elements into hierarchical
relations using hand-crafted heuristics. It explicitly considers
multi-column layouts and geometric cues such as indentation,
relative position, and spacing, but largely ignores richer
meta-information such as the actual text content of elements.
As a result, DocParser is effective on clean, well-structured
layouts but struggles on noisy scans and long documents where
semantic signals are crucial.

\paragraph{DSG.}
DSG~\cite{rausch2023dsgendtoenddocumentstructure} replaces heuristic rules with an
end-to-end neural relation predictor. It leverages a bidirectional
LSTM to model relations between layout elements, using visual
features extracted from an FPN backbone for image regions and
GloVe word embeddings for layout element types. Compared to
DocParser, DSG better captures local context among nearby
blocks, but it is still limited by its reliance on relatively shallow
sequence modeling and can degrade on highly irregular or
domain-shifted layouts.

\paragraph{DSPS.}
DSPS~\cite{10.1609/aaai.v37i2.25277} is the baseline introduced with the HRDoc
dataset. It employs a multi-modal encoder and a GRU decoder
for hierarchical organization. Textual embeddings of layout
elements are extracted separately and fused with geometric and
visual features inside the encoder. The decoder then predicts
parent--child relations in an autoregressive manner. This design
improves robustness by jointly modeling text and layout, but
the GRU-based decoding and local decision process make it
difficult to enforce a globally optimal tree, especially across
multiple pages.

\paragraph{DSHP-LLM.}
DSHP-LLM~\cite{multidocfusion} is an LLM-based DHP model that takes as input
a textualized representation of each document, including
block indices, layout types, and (optionally) truncated text.
A fine-tuned instruction-following LLM (e.g., Mistral-8B)
is prompted to output, for every block, the identifier of its
parent or a special ROOT symbol. This approach benefits
from strong long-context reasoning and flexible natural
language prompting, but remains sensitive to prompt
design, context-window limits, and instability in sequence
generation, especially for cross-page links and complex
layouts.

\paragraph{Qwen2.5-VL--DHP--SFT.}
Qwen2.5-VL--DHP--SFT adapts the DSHP-LLM idea to a
multi-modal LVLM backbone. It retains the decoder-style
sequence generation objective (predicting parent identifiers
from textualized blocks) while allowing the model to access
visual cues through image embeddings. In our experiments
it serves as a strong SFT-only baseline: Qwen2.5-VL--
DHP--SFT uses the same training data and prompts as
DSHP-LLM but does \emph{not} attach a biaffine head or
perform MST-based decoding. This highlights the
contribution of explicit dependency scoring and global
tree constraints in M3DocDep.

\paragraph{Training and evaluation protocol.}
For all DHP baselines, we follow the official train/test splits of DocHieNet, HRDH, and HRDS and adopt the hyperparameters recommended in the original papers. In particular, DocParser, DSG, and DSPS are run with their official public implementations and default settings, and DSHP-LLM and Qwen2.5-VL--DHP--SFT are trained with the same learning rates, batch sizes, and optimization schedules reported in their respective works. As in DocHieNet~\cite{dochienet} and MultiDocFusion~\cite{multidocfusion}, supervision and evaluation are defined at the block level: each annotated block is treated as a node with a single parent (or ROOT), and models are trained and evaluated by predicting the parent of every non-ROOT block. In Table~1 of the main paper, the tree-aware baselines and M3DocDep are evaluated on the same SharedDet blocks, where they consume the same Global Document Blocks produced by SharedDet. The general-purpose LVLM rows are included as reference baselines rather than part of this shared-block comparison. This isolates hierarchy recovery from differences in document parsing and OCR for the tree-aware methods.

\begin{table}[t]
\centering
\footnotesize
\renewcommand{\arraystretch}{1.05}
\setlength{\tabcolsep}{2pt}
\begin{tabular}{p{0.23\linewidth} >{\raggedright\arraybackslash}p{0.41\linewidth} >{\raggedright\arraybackslash}p{0.30\linewidth}}
\toprule
\textbf{Method} & \textbf{Key hyperparameters} & \textbf{Notes} \\
\midrule
Length & window=550 token & only text, no structure \\
\hline
Semantic & base encoder=E5 & sentence-level clustering \\
\hline
LumberChunker & backbone=Mistral-8B & topic-shift prompts \\
\hline
Perplexity & backbone=Mistral-8B, perplexity window tuned & Meta-Chunking \\
\hline
Structure-based & uses DP types & Layout based chunks \\
\hline
MultiDocFusion & uses DHP tree, max\_len=550 tok & tree-based chunks \\
\bottomrule
\end{tabular}
\caption{Typical configuration of chunking baselines.
Exact values per dataset are provided in the released configs.}
\label{tab:baseline_config}
\end{table}

\subsection{Chunking Baselines}
\label{sec:appendix_chunking_methods}

This subsection provides comprehensive descriptions of the
chunking methodologies compared against our proposed
tree-based structure-aware dependency chunking in
\textbf{M3DocDep}. Each chunking method is illustrated
with examples in Table~\ref{tab:chunking_examples}.

\paragraph{Length chunking \cite{Gong2020Recurrent}}
This method divides documents into chunks based on a fixed
token length limit. Each chunk is created uniformly, without
considering semantic or structural boundaries. While simple
and computationally efficient, it risks splitting important
contexts, leading to potential information loss and degraded
performance in retrieval and QA tasks.

\paragraph{Semantic chunking \cite{qu-etal-2025-semantic}}
Semantic chunking leverages encoder-based language
models to maintain semantic consistency. Chunks are
formed by grouping sentences based on semantic similarity
scores derived from language models (e.g., E5 embeddings).
Although effective in maintaining semantic coherence, it
tends to produce shorter, numerous chunks, potentially
impacting retrieval efficiency. Following prior work
\cite{hong-etal-2024-intelligent}, we employed the E5
model for consistency in our experiments.

\paragraph{LumberChunker \cite{duarte2024lumberchunker}}
LumberChunker employs Large Language Models (LLMs) to
dynamically partition documents by identifying topical
shifts between sentences or paragraphs. It effectively
captures the semantic independence of textual segments,
resulting in chunks of variable sizes optimized for dense
retrieval tasks. For experimental consistency across
LLM-based methods, we employed the Mistral-8B model as
the base model.

\paragraph{Perplexity chunking \cite{zhao2024metachunking}}
Based on the concept of Meta-Chunking, Perplexity
chunking identifies optimal chunk boundaries by analyzing
the perplexity distribution of sentences and paragraphs. It
dynamically merges or splits textual segments at a
fine-grained level, effectively balancing granularity and
computational efficiency. To ensure fairness among
LLM-based methods, we also used the Mistral-8B model for
these experiments.

\paragraph{Structure-based Chunking}
This approach partitions documents solely based on their
structural layouts, such as section headers, tables, and
figures. Similar methodologies have been explored in recent
works \cite{yepes2024financialreportchunkingeffective,
verma2025s2chunkinghybridframework}. In our experiments,
Structure-based Chunking served as a baseline to clearly
isolate and demonstrate the impact of stronger hierarchical
parsers. Specifically, chunks were created by ordering
structural elements obtained via DP (Document Parsing),
without explicitly considering hierarchical parent--child
relationships identified by DSHP-LLM or M3DocDep.
Segment types were included in the resulting chunks.

\paragraph{MultiDocFusion}
MultiDocFusion~\cite{multidocfusion} is a prior
hybrid multi-modal chunking pipeline that integrates
hierarchical document structure into the chunking process.
It utilizes a DSHP-LLM model (fine-tuned Mistral-8B)
identified in previous work to explicitly reconstruct section
hierarchies and then performs rule-based fusion of
hierarchy-aware segments into chunks. This significantly
enhances the semantic and structural coherence of chunks
compared to purely text- or layout-based baselines and
serves as a strong structure-aware chunking baseline in our
experiments. M3DocDep further improves upon this line of
work by replacing LLM-only hierarchy prediction with
LVLM-based dependency scoring and MST-based global
tree decoding, yielding more stable trees and more
boundary-faithful chunks.

\begin{table}[t]
\centering
\resizebox{\columnwidth}{!}{%
\begin{tabular}{lcc}
\toprule
 & \textbf{MultiDocFusion} & \textbf{M3DocDep} \\
\midrule
Hierarchy recovery & LLM-based hierarchical parsing & LVLM embeddings + MST global constraint \\
Visual handling & Absorbed into OCR text & Preserves table/figure crops + captions \\
Chunking signal & Structure $\rightarrow$ chunk & Multimodal tree recovery $\rightarrow$ indexing signal \\
\bottomrule
\end{tabular}%
}
\caption{Conceptual comparison between MultiDocFusion and M3DocDep. The main difference is not merely stronger supervision, but explicit multimodal dependency recovery with globally constrained tree decoding.}
\label{tab:supp_multidocfusion_comp}
\end{table}

\subsection{Chunking Configuration}

Table~\ref{tab:baseline_config} summarizes the hyperparameters and design choices for all chunking baselines used in our experiments. Each method follows its original formulation, but all chunkers operate on the same Global Document Blocks produced by SharedDet, use the same maximum chunk length (550 tokens), and are evaluated under the same corpus-level retrieval setting. This shared protocol is intended to isolate the effect of chunk construction rather than differences in parsing, chunk budget, or retrieval setup.

% ============================================================

\section{Implementation Details of M3DocDep}
\label{sec:impl_supp}

\begin{figure*}[t]
\centering

% -------- Figure 1: (a) + (b) top row --------
\begin{minipage}[t]{\textwidth}
  \centering
  \setlength{\tabcolsep}{2pt}
  \begin{tabular}{ccccc}
    \includegraphics[height=0.14\textheight,trim=40 60 40 70,clip]{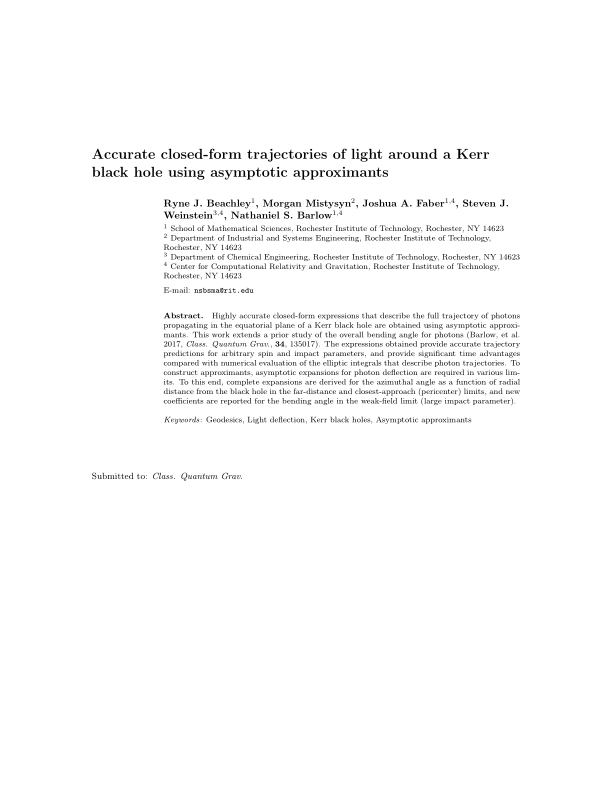} &
    \includegraphics[height=0.14\textheight,trim=40 60 40 70,clip]{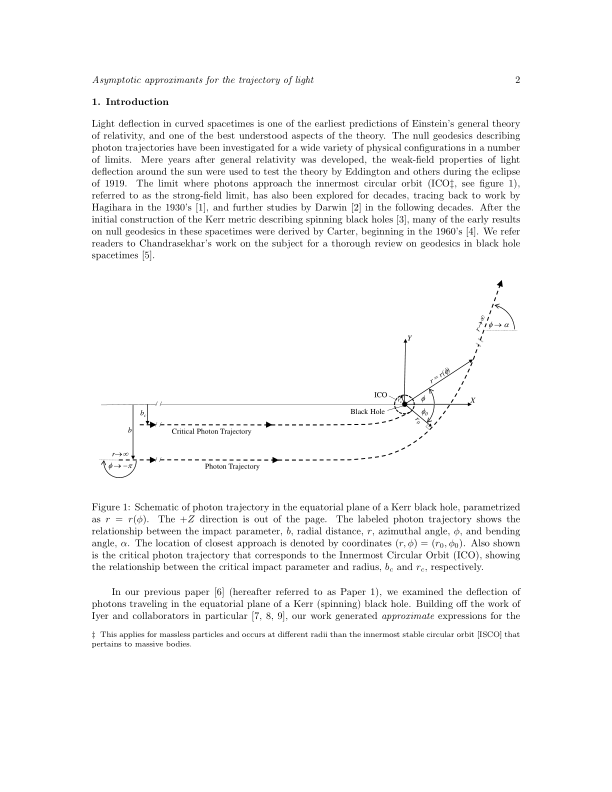} &
    \includegraphics[height=0.14\textheight,trim=40 60 40 70,clip]{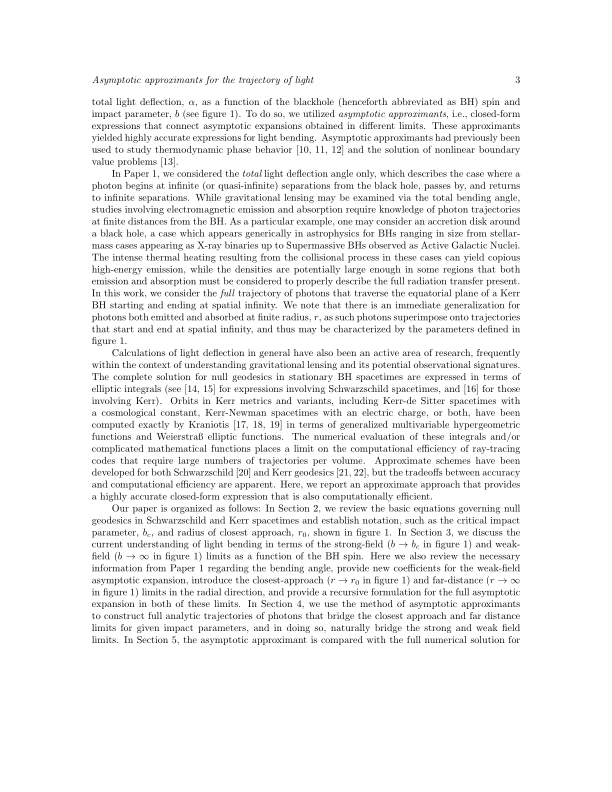} &
    \includegraphics[height=0.14\textheight,trim=40 60 40 70,clip]{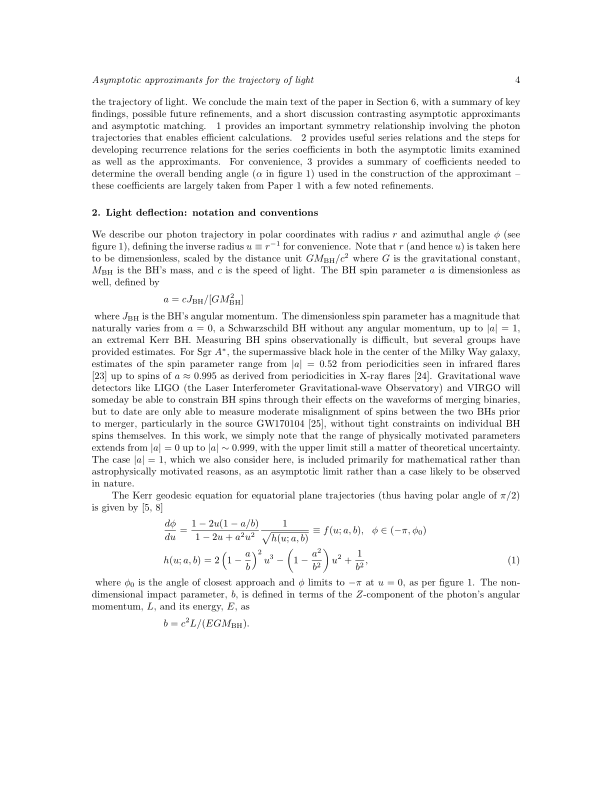} &
    \includegraphics[height=0.14\textheight,trim=40 60 40 70,clip]{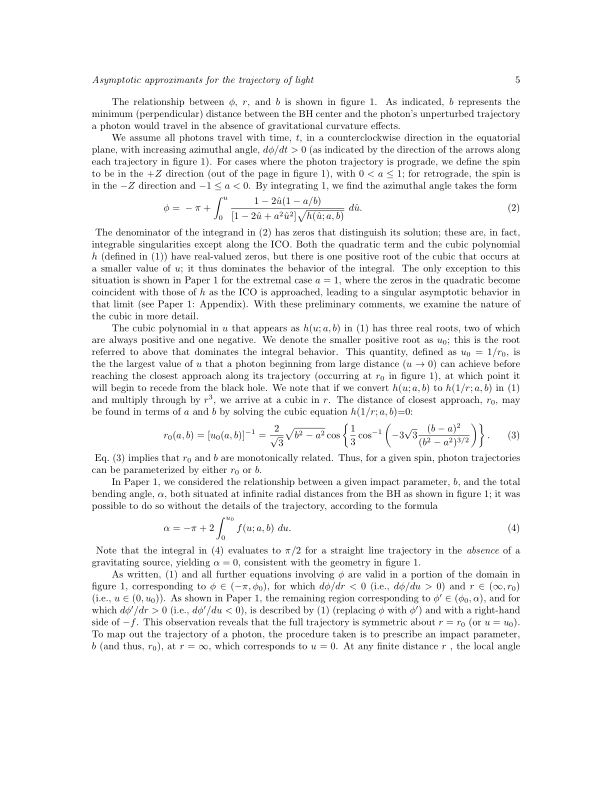}
  \end{tabular}\\[2pt]
  {\footnotesize (a) \textbf{Multi-page document} --- the original 5-page industrial guideline before any processing.}

  \medskip

  \begin{tabular}{ccccc}
    \includegraphics[height=0.15\textheight,trim=70 60 60 70,clip]{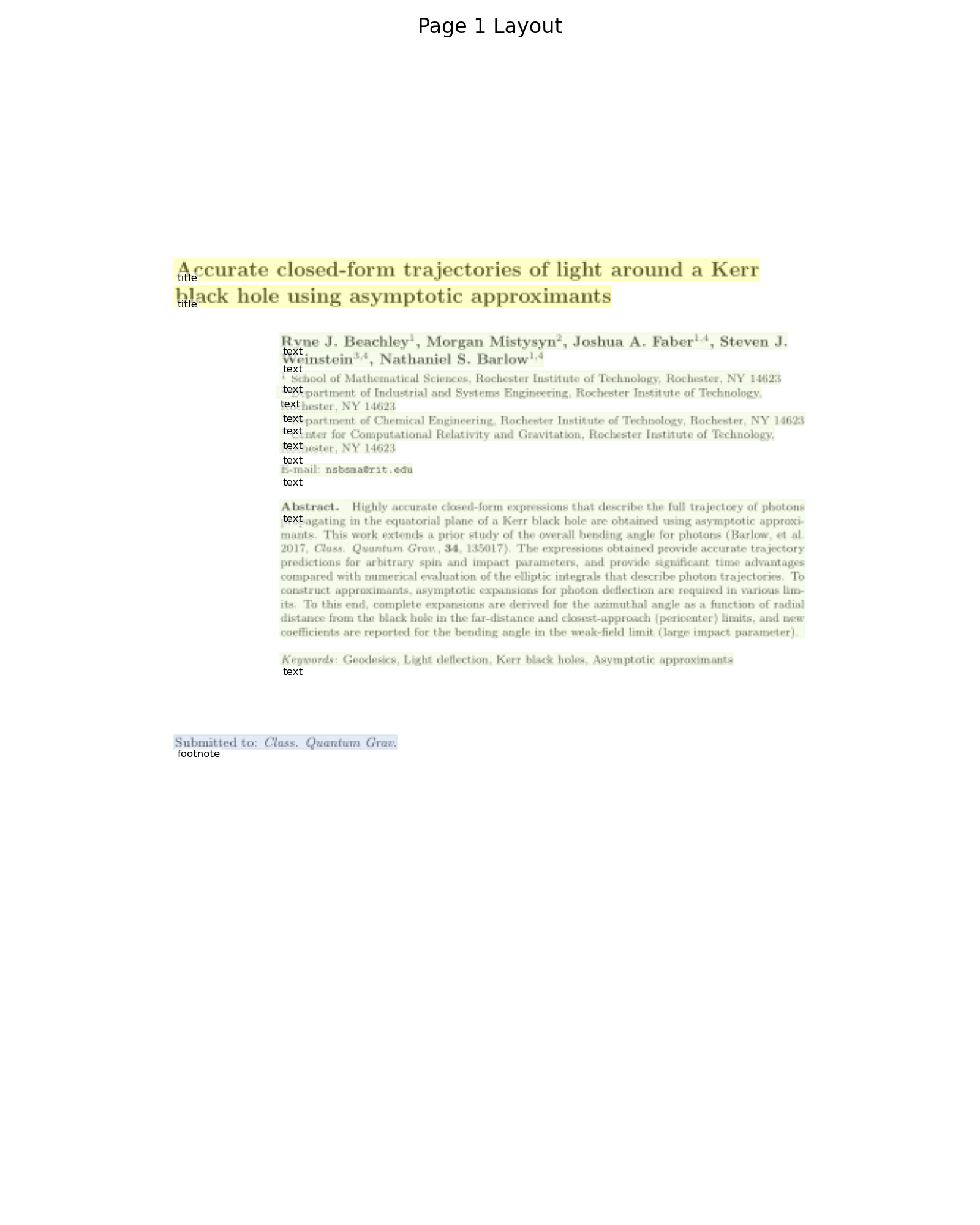} &
    \includegraphics[height=0.15\textheight,trim=70 60 60 70,clip]{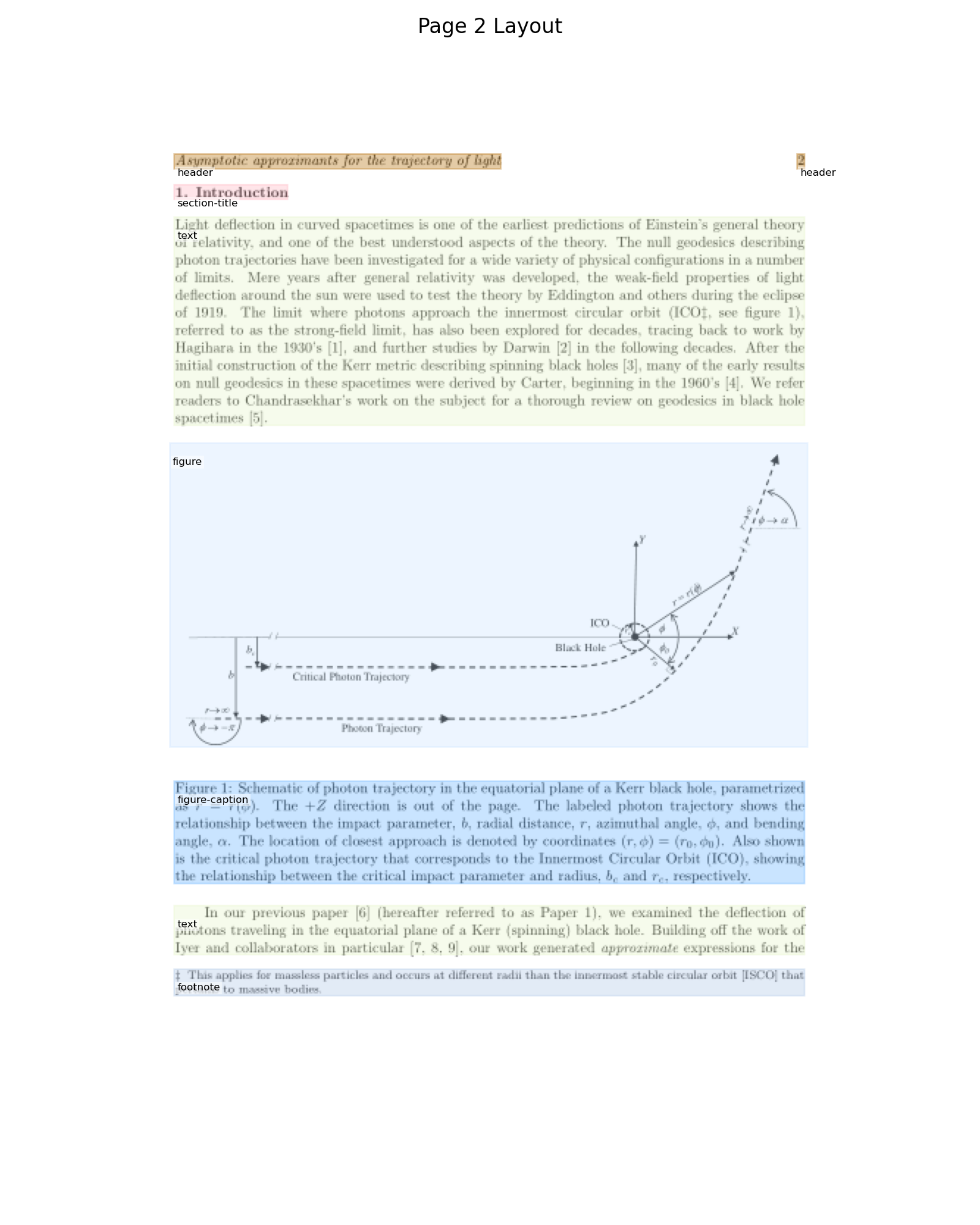} &
    \includegraphics[height=0.15\textheight,trim=70 60 60 70,clip]{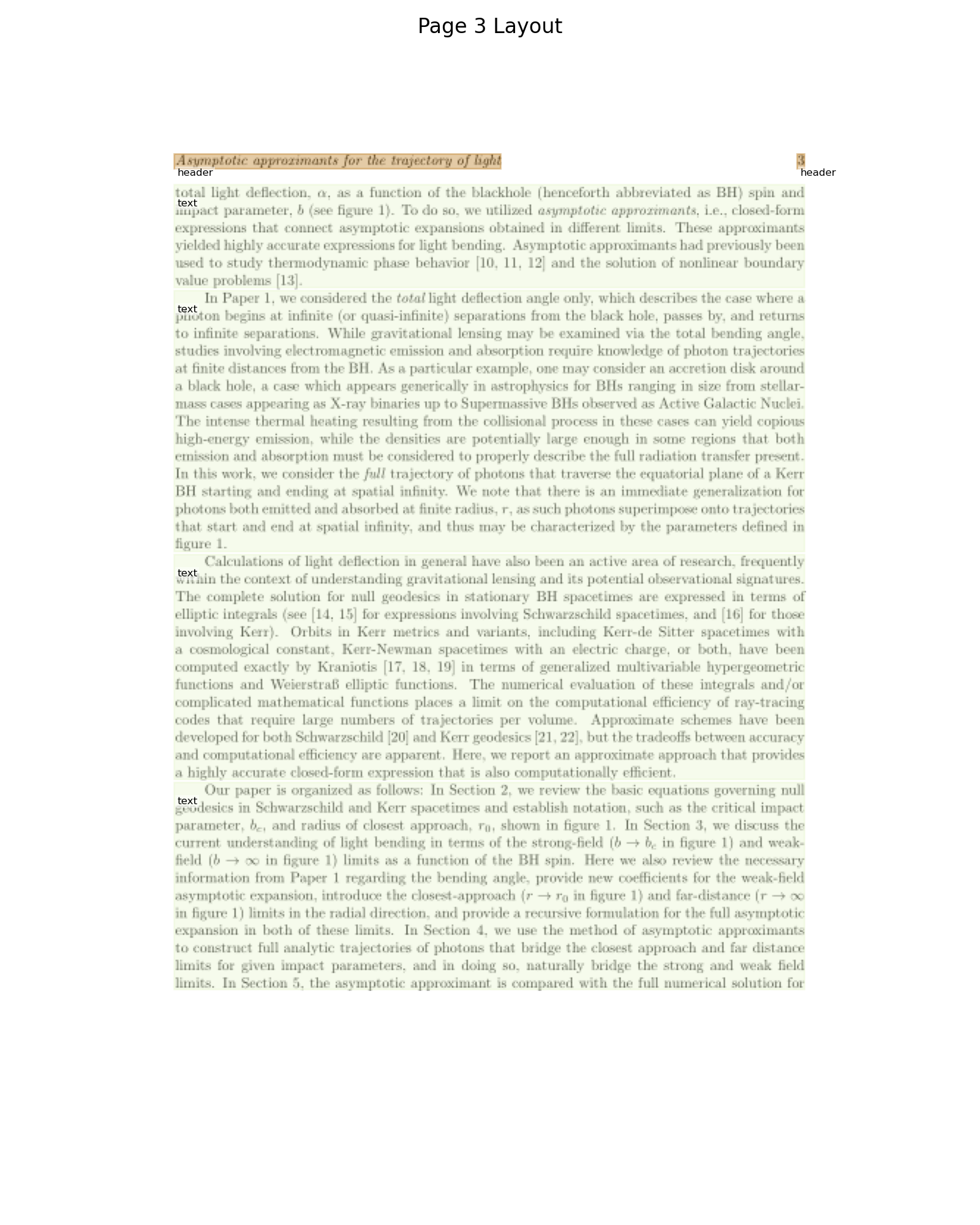} &
    \includegraphics[height=0.15\textheight,trim=70 60 60 70,clip]{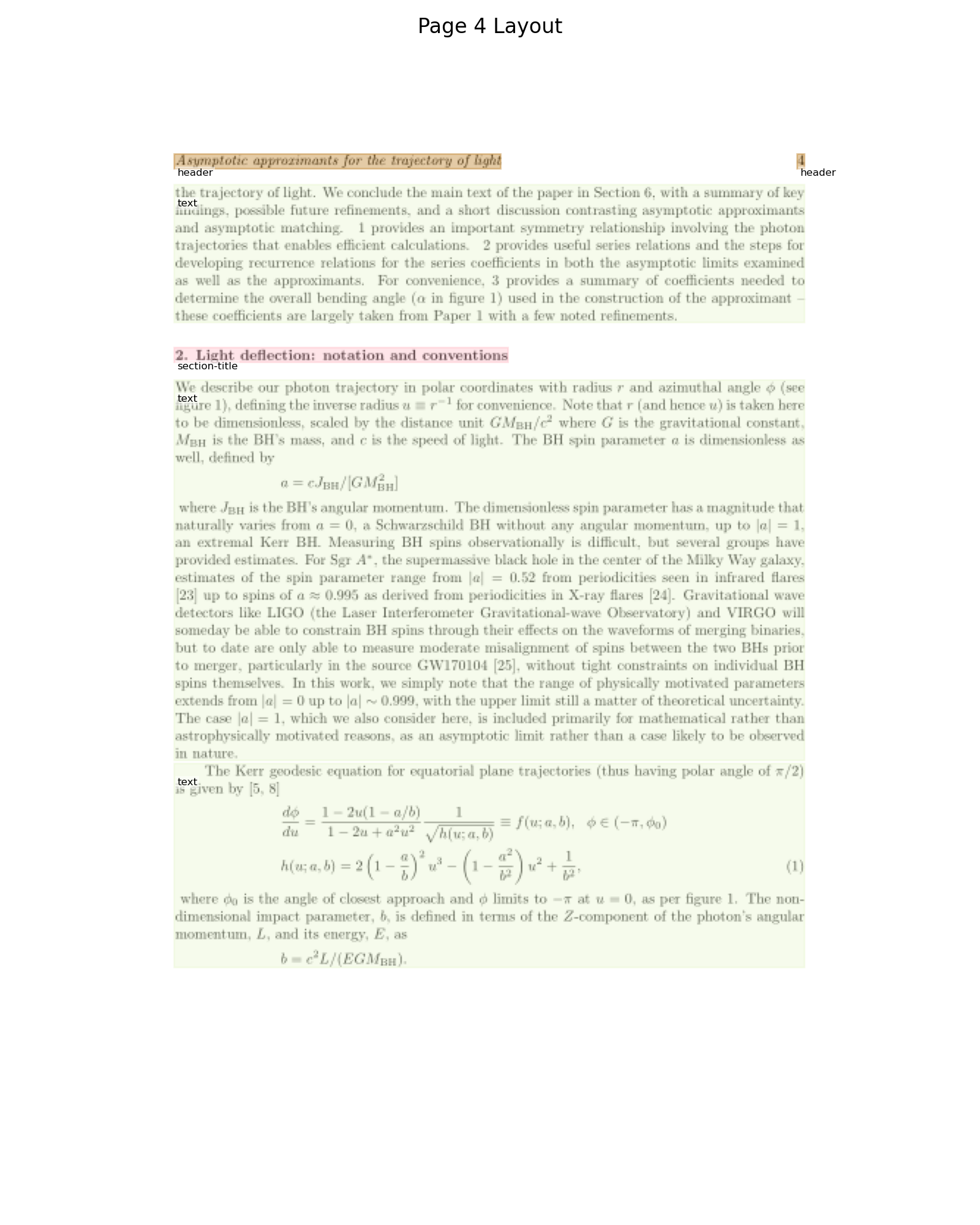} &
    \includegraphics[height=0.15\textheight,trim=70 60 60 70,clip]{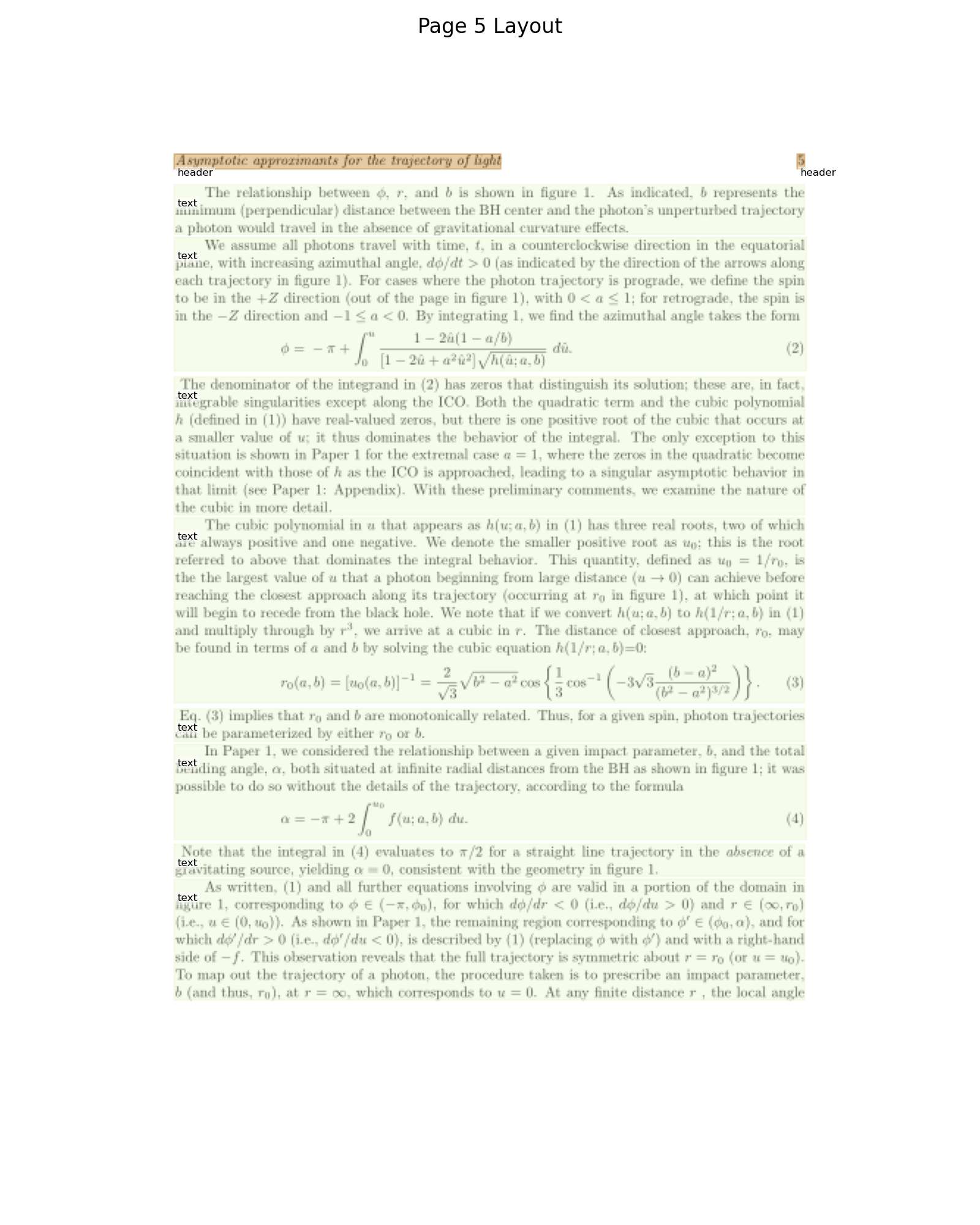}
  \end{tabular}\\[2pt]
  {\footnotesize (b) \textbf{SharedDet(DP + OCR)} --- page-level layout regions detected by DP and linked with OCR text to form Global Document Blocks.}
  \captionof{figure}{Structure-aware pipeline (part~1): (a) raw 5-page document and (b) DP + OCR outputs used to construct Global Document Blocks.}
  \label{fig:document_example_steps_ab_real}
\end{minipage}

\bigskip\hrule\bigskip

% -------- Figure 2: (c) + (d) bottom row --------
\begin{minipage}[t]{\textwidth}
  \centering
  \includegraphics[width=0.98\textwidth,height=0.25\textheight,keepaspectratio]{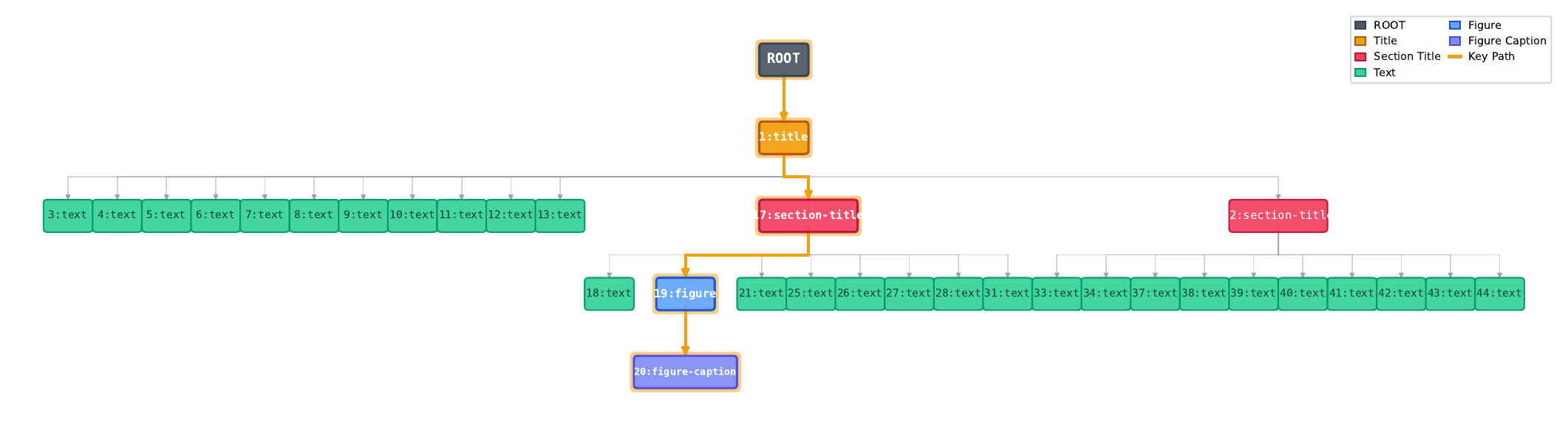}\\[2pt]
  {\footnotesize (c) \textbf{Global Document Dependency Tree} --- section headers, paragraphs, tables, and figures organized into a hierarchical tree with parent--child links.}

  \medskip

  \resizebox{!}{0.14\textheight}{%
    \begin{tikzpicture}[font=\scriptsize, x=1cm,y=1cm]
      \tikzset{
        chunkframe/.style={rounded corners,draw=blue!70!black,thick,minimum width=2.6cm,minimum height=3.0cm},
        chunksec/.style={fill=blue!10,text width=2.3cm,align=left,inner sep=2pt,font=\scriptsize\ttfamily},
        chunkmeta/.style={fill=yellow!10,text width=2.3cm,align=left,inner sep=2pt},
        chunkctx/.style={fill=green!8,text width=2.3cm,align=left,inner sep=2pt}
      }
      \node[chunkframe,anchor=west] (c1) at (0,0) {};
      \node[anchor=south,font=\scriptsize\bfseries] at ([yshift=2pt]c1.north) {Chunk 1};
      \node[chunksec,anchor=north west] (c1sec) at ([xshift=0.15cm,yshift=-0.15cm]c1.north west) {\# title};
      \node[chunkmeta,anchor=north west] (c1meta) at ([yshift=-0.1cm]c1sec.south west) {pages: 1};
      \node[chunkctx,anchor=north west] at ([yshift=-0.1cm]c1meta.south west) {context: text};
      \node[chunkframe,anchor=west] (c2) at (3.0,0) {};
      \node[anchor=south,font=\scriptsize\bfseries] at ([yshift=2pt]c2.north) {Chunk 2};
      \node[chunksec,anchor=north west] (c2sec) at ([xshift=0.15cm,yshift=-0.15cm]c2.north west) {\# title\\\#\# section-title 1};
      \node[chunkmeta,anchor=north west] (c2meta) at ([yshift=-0.1cm]c2sec.south west) {pages: 2};
      \node[chunkctx,anchor=north west] at ([yshift=-0.1cm]c2meta.south west) {context: text};
      \node[chunkframe,anchor=west] (c3) at (6.0,0) {};
      \node[anchor=south,font=\scriptsize\bfseries] at ([yshift=2pt]c3.north) {Chunk 3};
      \node[chunksec,anchor=north west] (c3sec) at ([xshift=0.15cm,yshift=-0.15cm]c3.north west) {\# title\\\#\# section-title 1};
      \node[chunkmeta,anchor=north west] (c3meta) at ([yshift=-0.1cm]c3sec.south west) {page: 2};
      \node[chunkctx,anchor=north west] at ([yshift=-0.1cm]c3meta.south west) {figure + caption};
      \node[chunkframe,anchor=west] (c4) at (9.0,0) {};
      \node[anchor=south,font=\scriptsize\bfseries] at ([yshift=2pt]c4.north) {Chunk 4};
      \node[chunksec,anchor=north west] (c4sec) at ([xshift=0.15cm,yshift=-0.15cm]c4.north west) {\# title\\\#\# section-title 1};
      \node[chunkmeta,anchor=north west] (c4meta) at ([yshift=-0.1cm]c4sec.south west) {pages: 2--4 (cross-page)};
      \node[chunkctx,anchor=north west] at ([yshift=-0.1cm]c4meta.south west) {context: text};
      \node[chunkframe,anchor=west] (c5) at (12.0,0) {};
      \node[anchor=south,font=\scriptsize\bfseries] at ([yshift=2pt]c5.north) {Chunk 5};
      \node[chunksec,anchor=north west] (c5sec) at ([xshift=0.15cm,yshift=-0.15cm]c5.north west) {\# title\\\#\# section-title 2};
      \node[chunkmeta,anchor=north west] (c5meta) at ([yshift=-0.1cm]c5sec.south west) {pages: 4--5 (cross-page)};
      \node[chunkctx,anchor=north west] at ([yshift=-0.1cm]c5meta.south west) {context: text};
    \end{tikzpicture}%
  }\\[2pt]
  {\footnotesize (d) \textbf{Tree-based Structure-Aware Dependency Chunking} --- five example chunks. \textcolor{blue!70!black}{Blue}: section path, \textcolor{yellow!60!black}{yellow}: metadata, \textcolor{green!50!black}{green}: context type.}
  \captionof{figure}{Structure-aware pipeline (part~2): (c) Global Document Dependency Tree reconstruction and (d) Tree-based Structure-Aware Dependency Chunking.}
  \label{fig:document_example_steps_cd_real}
\end{minipage}

\end{figure*}

This section expands on Sec.~3 of the main paper by detailing each component of M3DocDep and its training configuration.
Figures~\ref{fig:document_example_steps_ab_real} and~\ref{fig:document_example_steps_cd_real} also include corresponding real-world examples to illustrate each step of the pipeline.

\subsection{SharedDet (DP+OCR)}
We reuse the SharedDet pipeline defined in Sec.~\ref{sec:preprocess_supp} to produce
\emph{Global Document Blocks}. In the rest of this section we focus on how these
blocks are consumed by M3DocDep (embedding, parsing, and chunking).

\subsection{LVLM-based Multi-modal Block Embedding}

M3DocDep uses frozen LVLM encoders to extract page-level visual tokens and aggregates
them into block embeddings via SoftROI (see the list of backbones in Sec.~\ref{sec:lvlm_rag_supp}).

\paragraph{Page Multi-modal Tokens.}
Each page image is fed into a frozen LVLM. From the last
decoder layer we extract hidden states at positions
corresponding to visual tokens. Using token-grid metadata,
we map each token to 2D coordinates in the global
document frame $[0,1]^2$, yielding a set of tokens
$\{z_p\}$ per page with coordinates $(u_p, v_p)$.

\paragraph{SoftROI pooling.}
For each block $i$ with normalized box $\mathrm{bbox}_i$, the
SoftROI Embedder collects tokens whose coordinates lie
inside the box and assigns them boundary-aware weights:
\[
w_p \propto [u_p(1-u_p)]^\alpha [v_p(1-v_p)]^\alpha,
\quad \tilde{w}_p = \frac{w_p}{\sum_{q \in \text{ROI}_i} w_q},
\]
where $\alpha$ controls boundary sharpness. The SoftROI
Multi-modal Block Embedding is
\[
e_i = \sum_{p \in \text{ROI}_i} \tilde{w}_p z_p.
\]
Compared to uniform pooling, this applies a spatially-aware
weighting that downweights tokens near box edges and corners,
making the embedding more robust to box jitter and imperfect
detections.

\paragraph{Type-aware embeddings.}
Block types (title, section header, paragraph, table, figure,
caption, other) are mapped to a small embedding table
$\tau_i$. We concatenate SoftROI embeddings and type
embeddings, $x_i = [e_i; \tau_i]$, and pass them through a
two-layer MLP to obtain hidden representations $h_i$ used
for dependency scoring.

\subsection{Global Document Dependency Parsing}

We adopt a biaffine dependency scoring head to score
parent--child relations between blocks and recover a global
document tree.

\paragraph{Candidate parent selection.}
For each child block $v$, we construct a small candidate
parent set $P(v)$ by:
\begin{itemize}[leftmargin=*]
  \item prioritizing title and section-header blocks;
  \item allowing ``upward'' links on the same page within a
        vertical tolerance;
  \item allowing cross-page parents only within the most
        recent $M$ pages;
  \item discarding implausible parents based on type
        constraints (e.g., paragraphs rarely parent top-level
        titles).
\end{itemize}
We keep at most $K$ candidates per child based on a
header- and distance-based heuristic; $(M, K)$ are tuned on
a validation subset and fixed across datasets.

\paragraph{Biaffine scoring.}
Given hidden states $h_i$, we score edges $u \rightarrow v$
using a biaffine function with geometric features:
\[
s(u \rightarrow v) = [h_u;1]^\top U [h_v;1] +
w_{\text{geo}}^\top \delta_g(u,v),
\]
where $\delta_g(u,v)$ includes relative offsets, size ratios,
page-distance, and overlap indicators. The virtual root
score is defined as $s(\mathrm{ROOT}\rightarrow v) =
r^\top h_v + b_r$.

For each child $v$ we normalize scores over
$P(v) \cup \{\mathrm{ROOT}\}$ with a $(K+1)$ child-softmax
and minimize cross-entropy against the ground-truth parent.

\paragraph{MST-based global tree decoder.}
At inference time, we treat edge scores as weights and feed
them into the Chu--Liu/Edmonds algorithm~\cite{chuliu1965,edmonds1967} to obtain the
maximum spanning arborescence, enforcing single-root,
single-parent, and acyclicity constraints. We also measure a
local argmax baseline that chooses the best-scoring parent
per child without global constraints.

\subsection{Tree-based Structure-Aware Dependency Chunking}

Given the Global Document Dependency Tree, M3DocDep first applies a DFS-based dependency chunking procedure over the tree, which is summarized in Algorithm~\ref{alg:appendix_dfs}. Concrete examples of the resulting chunks are illustrated in Figure~\ref{fig:document_example_steps_cd_real}.

\begin{algorithm}[h]
\footnotesize
\caption{Tree-based Structure-Aware Dependency Chunking}
\label{alg:appendix_dfs}
\begin{algorithmic}[1]
\Require Global dependency tree $G=(V,E)$, max chunk length $\texttt{max\_len}$
\Function{BuildChunks}{$\texttt{root}$}
  \State $\texttt{chunks} \gets [\,]$
  \State \Call{DFSChunk}{$\texttt{root}$, [], "", \texttt{chunks}}
  \State \Return \texttt{chunks}
\EndFunction
\Function{DFSChunk}{$v$, path, buffer, chunks}
  \State path $\gets$ path $+\,[v]$
  \State buffer $\gets$ buffer $+\,$text($v$)
  \If{length(buffer) $>$ \texttt{max\_len}}
    \State split buffer into one or more chunks and append to \texttt{chunks}
    \State reset buffer to last partial chunk
  \EndIf
  \For{child $\in$ children($v$) in doc order}
    \State \Call{DFSChunk}{child, path, buffer, chunks}
  \EndFor
\EndFunction
\end{algorithmic}
\end{algorithm}

\paragraph{Section subtree grouping.}
We mark title and section-header blocks as section roots and
perform DFS from each root to collect descendant blocks
into section subtrees. Blocks in the same subtree are merged
across page boundaries, so a chunk can span multiple pages
when a section continues.

\paragraph{Figure/table--caption binding.}
For blocks labeled as figures or tables, we exploit tree
structure: if a figure/table and a caption form a parent--child
(or closely related) relation, they are forced into the same
chunk. When no explicit edge exists, we fall back to spatial
proximity and reading order heuristics on the page.

\paragraph{Section path and metadata.}
Each chunk is annotated with:
(i) section path from root to governing header,
(ii) page range, and
(iii) constituent block IDs and layout types.
This metadata is stored in the retrieval index and used
during RAG to present structure-aware context to the LVLM.

\paragraph{Logical consistency.}
Because the dependency parser decodes a maximum spanning tree, every non-root block receives exactly one parent and the recovered structure is globally acyclic. The chunker therefore operates on a valid tree rather than on a set of independently chosen local links, which makes figure--caption binding and cross-page section grouping more stable.

\paragraph{Granularity control.}
Chunk granularity is adjusted deterministically on the recovered tree by changing the maximum chunk length and the cut policy. In practice, this allows section-level, paragraph-level, or finer chunking without retraining the dependency parser: coarse settings keep larger subtrees intact, while finer settings cut earlier along long paths or large sibling groups.

\subsection{Training Hyperparameters}

We train only the dependency head while keeping the LVLM encoders frozen, as described in Secs.~D.1--D.4 above. The main hyperparameters are summarized in Table~\ref{tab:supp_hparams}.

\begin{table}[t]
\centering
\renewcommand{\arraystretch}{1.1}
\setlength{\tabcolsep}{3pt}

\resizebox{\linewidth}{!}{
\begin{tabular}{
  >{\raggedright\arraybackslash}p{\compw}%
  >{\raggedright\arraybackslash}p{\hpw}}
\toprule
\textbf{Component} & \textbf{Hyperparameters (default / range)} \\
\midrule
\multirow[t]{8}{*}{Dependency head}
 & Learning rate: $1\!\times\!10^{-5}$--$5\!\times\!10^{-5}$ \\
 & Optimizer: Adam / AdamW \\
 & Batch size: 8--16 docs/GPU \\
 & Epochs: 3--5 (early stop) \\
 & Dropout: 0.0--0.1 (MLP layers) \\
 & Weight decay: $0$--$10^{-2}$ \\
 & Parent window $M$: recent pages (tuned on val) \\
 & Candidate top-$K$: $\{8,16,32\}$ (ablations) \\
\bottomrule
\end{tabular}
} 
\caption{Summary of key hyperparameters used to train the M3DocDep dependency
head. Ranges denote grid/line searches; defaults follow the settings used
for the main tables.}
\label{tab:supp_hparams}
\end{table}

% ============================================================

\subsection{Training and Inference Environment}
\label{sec:env_supp}

All experiments are conducted on a GPU cluster equipped with 8 NVIDIA A100-SXM4-80GB GPUs (80 GB VRAM each), 64-core CPUs, and 1 TB of system RAM. We implement M3DocDep in PyTorch (v2.2) with CUDA (v12.9), and use FAISS (v1.8) to build dense retrieval indices. Training the dependency head on HRDH for 3 epochs takes about 6 hours on a single A100 GPU, including data loading and evaluation. For the per-page runtime numbers in Table~\ref{tab:supp_runtime}, we use a single A100 (80GB) and observe 27 GB peak GPU memory during end-to-end indexing. Corpus-level indexing for DUDE, MP-DocVQA, CUAD, and MOAMOB requires approximately 1--3 hours per corpus, while full QA evaluation takes an additional 2--4 hours per corpus.

% ============================================================
\section{LVLM and RAG Setup}
\label{sec:lvlm_rag_supp}

This section details the LVLM backbones, retrieval pipeline, and prompting strategy used across all experiments. Table~\ref{tab:supp_rag_config} summarizes the full RAG configuration.

\begin{table}[t]
\centering
\renewcommand{\arraystretch}{1.1}
\setlength{\tabcolsep}{3pt}

\resizebox{\linewidth}{!}{%
\begin{tabular}{
  >{\raggedright\arraybackslash}p{\compw}%
  >{\raggedright\arraybackslash}p{\hpw}}
\toprule
\textbf{Setting} & \textbf{Configuration} \\
\midrule
Corpus & DUDE, MP-DocVQA, CUAD, MOAMOB \\
Index granularity & Chunk-level, corpus-level index \\
Retriever (dense) & BGE, E5, MM-Embed \\
Retriever (sparse) & BM25 \\
$k_{\text{ret}}$ & $\{1,2,3,4\}$ \\
Reader LVLMs & LLaVA-OneVision-1.5, InternVL-3.5, Qwen2.5-VL \\
Metrics (retrieval) & Recall, Precision, nDCG \\
Metrics (QA) & ANLS, ROUGE-L, METEOR \\
Prompting & Instruction + query + serialized chunk list \\
Decoding & Fixed temperature/top-$p$ per LVLM \\
\bottomrule
\end{tabular}
}
\caption{Summary of RAG configuration used in our experiments.}
\label{tab:supp_rag_config}
\end{table}

\subsection{LVLM Backbone Configurations}

We use three open-source LVLM backbones:
LLaVA-OneVision-1.5~\cite{an2025llavaonevision15fullyopenframework},
InternVL-3.5~\cite{wang2025internvl35advancingopensourcemultimodal},
and Qwen2.5-VL~\cite{bai2025qwen25vltechnicalreport}. For all of them:

\begin{itemize}[leftmargin=*]
  \item Input pages are rendered at a shared resolution and
        fed either one page per call or in small page batches,
        depending on the model's context window.
  \item We fix the maximum number of images per call and
        slice long documents across multiple calls if needed.
  \item Decoding parameters (temperature, top-$p$,
        \texttt{max\_new\_tokens}) are kept identical across
        all chunking methods within each experiment.
\end{itemize}

For GPT-5 and other closed LVLM baselines, we record and report the provider, model identifier, API mode/endpoint, access date, prompt template, and decoding parameters (temperature, top-$p$, and output-token limit) used in each experiment. These baselines are used only for comparison; M3DocDep itself does not rely on closed models.

\subsection{Retrieval Pipeline}

All chunking methods share the same retrieval backbone.

\paragraph{Dense and sparse retrieval.}
We use dense embedding models such as
BGE~\cite{chen2024bgem3embeddingmultilingualmultifunctionality},
E5~\cite{wang2024multilinguale5textembeddings}, and
MM-Embed~\cite{lin2025mmembeduniversalmultimodalretrieval}
to obtain chunk-level representations, storing them in a
FAISS-based ANN index. For sparse retrieval, we use
BM25~\cite{10.1561/1500000019} over chunk texts.

\paragraph{Corpus-level top-$k_{\text{ret}}$.}
For each query we retrieve the top
$k_{\text{ret}} \in \{1,2,3,4\}$ chunks from the corpus-level
index. Unless otherwise specified, $k_{\text{ret}}=4$ is used in
the main tables, while extended experiments sweep over
$k_{\text{ret}}$ to analyze sensitivity.

\paragraph{Chunk serialization and reader input.}
Each chunk is serialized with a shared schema consisting of section path, page range, block-type markers, and OCR/caption text; fields unavailable to a given chunker are left blank or omitted. For figure/table chunks, the associated caption is kept in the same serialized unit so that retrieval preserves the figure--caption relation. Text-only retrievers (BGE, E5, BM25) operate on the shared serialized text, while MM-Embed additionally receives the associated figure/table crops when present. For multimodal readers, the corresponding figure/table crops are likewise passed alongside the serialized text when available. This shared serialization is used across chunking methods so that comparisons reflect chunk quality rather than reader-side formatting differences.

\begin{figure}[t]
  \centering
  \begin{tcolorbox}[
    title={DHP Prompt},
    colback=white,
    colframe=black,
    boxrule=0.4pt,
    arc=2pt,
    left=4pt,
    right=4pt,
    top=4pt,
    bottom=4pt,
    fonttitle=\bfseries
  ]
  \small
  You are a document hierarchy parser. Using the provided page images
  and the full list of layout blocks, predict the parent for \emph{every}
  block in the document. Your goal is to recover the global dependency
  tree.\\[4pt]

  Use only the information given in the input. Do not invent new block
  ids or additional keys. Avoid self-parenting, cycles, and parents that
  come \emph{after} the child in reading order (future parents).\\[6pt]

  \texttt{[Input]} A single JSON object with fields
  \texttt{"instruction"}, \texttt{"pages"}, and \texttt{"blocks"}.
  The \texttt{"pages"} field stores page sizes, and each entry in
  \texttt{"blocks"} contains the block id, page index, label,
  bounding box, and truncated text.\\[4pt]

  \texttt{[Output]} A JSON array of objects of the form
  \texttt{\{"id": "<BLOCK\_ID>", "parent": <PARENT\_ID or null>\}},
  one for each input block, in the same order as \texttt{"blocks"}.
  For any block where no valid parent can be assigned under the above
  constraints, set \texttt{"parent"} to \texttt{null}. Do not output
  anything other than this JSON array.
  \end{tcolorbox}
  \caption{Prompt used for VLM-only Document Hierarchical Parsing (DHP).}
  \label{fig:dhp_prompt}
\end{figure}

\subsection{DHP and QA Prompt}
\label{sec:prompt_templates}

We design two instruction-style prompts for our LVLM-based
components: (i) a VLM-only Document Hierarchical Parsing
(DHP) prompt that predicts a parent for every layout block,
and (ii) a RAG-style QA prompt that answers questions from
retrieved chunks. Both prompts are intentionally lightweight
and model-agnostic so that the same templates can be reused
across different LVLM backbones and datasets.

\paragraph{LVLM-based DHP.}
For the LVLM based DHP setting, the LVLM receives page images
together with the full list of detected layout blocks and is
asked to recover the global dependency tree by predicting a
single parent for each block. The prompt encodes page sizes
and block attributes (id, page, label, bounding box, truncated
text) and constrains the model to output a JSON-only list of
(\texttt{id}, \texttt{parent}) pairs that forms an acyclic tree
consistent with reading order. The exact DHP prompt template
is shown in Fig.~\ref{fig:dhp_prompt}.

\paragraph{RAG QA (LVLM read and generate).}
For downstream QA, the LVLM is given a natural-language question together with a small set of retrieved chunks, each tagged with its section path and page range and serialized under the shared chunk schema described above. The prompt asks the model to answer strictly based on these chunks and to explicitly abstain when the answer is not supported by the context, preventing hallucination and making RAG behavior easier to analyze. The QA prompt template is shown in Fig.~\ref{fig:qa_prompt}.

\begin{figure}[t]
  \centering
  \begin{tcolorbox}[
    title={QA Prompt},
    colback=white,
    colframe=black,
    boxrule=0.4pt,
    arc=2pt,
    left=4pt,
    right=4pt,
    top=4pt,
    bottom=4pt,
    fonttitle=\bfseries
  ]
  \small
  You are given a question about one or more technical documents.
  Use only the provided context chunks to answer the question.
  If the answer cannot be inferred from the context, say ``I don't know''.\\[4pt]

  \texttt{[Question]} \texttt{<QUESTION\_TEXT>}\\[2pt]
  \texttt{[Context chunks]}\\
  \texttt{<CHUNK\_1>}\\
  \texttt{<CHUNK\_2>}\\
  \texttt{<CHUNK\_3>}\\
  \texttt{<CHUNK\_4>}
  %\texttt{$\dots$}
  \end{tcolorbox}
  \caption{Prompt used for LVLM-based RAG QA over retrieved chunks.}
  \label{fig:qa_prompt}
\end{figure}

% ============================================================
\section{Extended Quantitative Analyses}
\label{sec:extra_results}

\subsection{Robustness Across Upstream Modules, Retrieval Backbones, and LVLMs}

Table~\ref{tab:supp_dp_ocr_swaps} extends the robustness analyses with stronger SharedDet backbones, Table~\ref{tab:supp_retriever_swap} adds retriever swaps, and Table~\ref{tab:supp_lvlm_swap} summarizes LVLM substitution results. The overall pattern is consistent: stronger upstream modules improve absolute performance for all structure-aware chunkers, while M3DocDep remains the best-performing method.

\begin{table*}[t]
\centering
\scriptsize
\setlength{\tabcolsep}{4pt}
\renewcommand{\arraystretch}{1.1}
\begin{tabular}{lccccc|ccccc}
\toprule
\multirow{2}{*}{\textbf{Method}} & \multicolumn{5}{c|}{\textbf{DP backbone (nDCG)}} & \multicolumn{5}{c}{\textbf{OCR backbone (nDCG)}} \\
\cmidrule(lr){2-6}\cmidrule(lr){7-11}
 & DETR & DiT & VGT & MinerU2.5 & DocLayout & EasyOCR & Tesseract & TrOCR & PaddleOCR & DotsOCR \\
\midrule
Structure-based & 0.4396 & 0.4171 & 0.4269 & 0.4516 & 0.4345 & 0.5194 & 0.4650 & 0.2993 & 0.5248 & 0.5384 \\
MultiDocFusion & 0.5014 & 0.4976 & 0.5061 & 0.5119 & 0.5047 & 0.5681 & 0.5068 & 0.4097 & 0.5742 & 0.5896 \\
\textbf{M3DocDep} & \textbf{0.5239} & \textbf{0.5127} & \textbf{0.5382} & \textbf{0.5532} & \textbf{0.5325} & \textbf{0.5914} & \textbf{0.5279} & \textbf{0.4235} & \textbf{0.5978} & \textbf{0.6136} \\
\bottomrule
\end{tabular}
\caption{Robustness across stronger SharedDet backbones. Values are macro-averaged nDCG over DUDE, MP-DocVQA, CUAD, and MOAMOB with top-$k \in \{1,2,3,4\}$. Stronger DP/OCR modules raise absolute performance, while M3DocDep remains best in every setting.}
\label{tab:supp_dp_ocr_swaps}
\end{table*}

\begin{table}[t]
\centering
\resizebox{0.92\columnwidth}{!}{%
\begin{tabular}{lccccc}
\toprule
\textbf{Chunking Method} & BGE & E5 & BM25 & MM-Embed & Avg \\
\midrule
Length chunking          & 0.4834 & 0.4715 & 0.4764 & 0.4864 & 0.4793 \\
Semantic chunking        & 0.3114 & 0.3378 & 0.1825 & 0.2906 & 0.2804 \\
LumberChunker            & 0.4708 & 0.4319 & 0.4539 & 0.4632 & 0.4549 \\
Perplexity chunking      & 0.4715 & 0.4318 & 0.4495 & 0.4647 & 0.4542 \\
Structure-based chunking & 0.4679 & 0.4040 & 0.4118 & 0.4591 & 0.4357 \\
MultiDocFusion           & 0.5213 & 0.4884 & 0.5085 & 0.5283 & 0.5116 \\
\textbf{M3DocDep}        & \textbf{0.5523} & \textbf{0.5014} & \textbf{0.5321} & \textbf{0.5654} & \textbf{0.5378} \\
\bottomrule
\end{tabular}%
}
\caption{Retrieval-backbone robustness. Values are macro-averaged nDCG over DUDE, MP-DocVQA, CUAD, and MOAMOB with top-$k \in \{1,2,3,4\}$. The benefit of M3DocDep is consistent across sparse, dense, and multimodal retrievers.}
\label{tab:supp_retriever_swap}
\end{table}

\begin{table}[t]
\centering
\resizebox{0.82\columnwidth}{!}{%
\begin{tabular}{lccc}
\toprule
\textbf{M3DocDep LVLM backbone} & Qwen2.5-VL & InternVL-3.5 & LLaVA-OneVision-1.5 \\
\midrule
DocHieNet parent F1 & 76.01 & 75.71 & 74.07 \\
\bottomrule
\end{tabular}%
}
\caption{LVLM swap for multimodal block embeddings under the shared DocHieNet DHP protocol. Performance remains stable across three open LVLM backbones.}
\label{tab:supp_lvlm_swap}
\end{table}

\subsection{Fairness Control for Structural Metadata}

To disentangle the effect of improved chunk boundaries from that of added metadata, we perform a pairwise fairness control in which section-path and page-range fields are removed from both MultiDocFusion and M3DocDep during indexing and reader input. Under this no-metadata control, M3DocDep still retains a 2.3\% nDCG advantage over MultiDocFusion. This confirms that the gain is not explained solely by metadata injection: better dependency recovery and boundary formation remain the primary source of improvement.

\subsection{Full Ablation on Dependency Recovery}

\begin{table*}[t]
\centering
\scriptsize
\renewcommand{\arraystretch}{1.15}
\setlength{\tabcolsep}{2.5pt}
\begin{tabularx}{0.84\textwidth}{@{} >{\raggedright\arraybackslash}X *{3}{cc} @{} }
\toprule
\textbf{Variant (Ablation)} &
\multicolumn{2}{c}{\textbf{HRDS}} &
\multicolumn{2}{c}{\textbf{HRDH}} &
\multicolumn{2}{c}{\textbf{DocHieNet}} \\
\cmidrule(lr){2-3}\cmidrule(lr){4-5}\cmidrule(lr){6-7}
& \textbf{F1} & \textbf{STEDS} & \textbf{F1} & \textbf{STEDS} & \textbf{F1} & \textbf{STEDS} \\
\midrule
\textbf{Full (SharedDet)} & \textbf{82.87} & \textbf{76.52} & \textbf{77.75} & \textbf{71.65} & \textbf{76.01} & \textbf{70.83} \\
\midrule
\emph{SoftROI} $\rightarrow$ \emph{uniform ROI pooling}
  & 81.64 (-1.23) & 74.85 (-1.67)
  & 76.81 (-0.94) & 70.13 (-1.52)
  & 74.43 (-1.58) & 68.92 (-1.91) \\
MST (global tree) $\rightarrow$ \emph{local argmax}
  & 78.31 (-4.56) & 70.59 (-5.93)
  & 71.92 (-5.83) & 64.19 (-7.46)
  & 70.82 (-5.19) & 64.12 (-6.71) \\
\emph{no header-centric parent prior}
  & 81.25 (-1.62) & 74.49 (-2.03)
  & 75.26 (-2.49) & 68.48 (-3.17)
  & 74.13 (-1.88) & 68.37 (-2.46) \\
candidate top-$k$ pruning: $k{=}8$
  & 81.61 (-1.26) & 74.79 (-1.73)
  & 76.12 (-1.63) & 69.47 (-2.18)
  & 74.55 (-1.46) & 68.89 (-1.94) \\
candidate top-$k$ pruning: $k{=}16$
  & 82.68 (-0.19) & 76.25 (-0.27)
  & 77.42 (-0.33) & 71.24 (-0.41)
  & 75.79 (-0.22) & 70.48 (-0.35) \\
candidate top-$k$ pruning: $k{=}32$
  & 82.51 (-0.36) & 76.04 (-0.48)
  & 77.28 (-0.47) & 71.06 (-0.59)
  & 75.63 (-0.38) & 70.21 (-0.62) \\
\emph{disallow cross-page edges}
  & 77.53 (-5.34) & 69.65 (-6.87)
  & 68.83 (-8.92) & 60.39 (-11.26)
  & 68.83 (-7.18) & 61.19 (-9.64) \\
\bottomrule
\end{tabularx}
\caption{Full per-dataset ablation on hierarchy and dependency reconstruction. Each cell is in the form score ($\Delta$), where $\Delta$ denotes the change relative to Full (SharedDet).}
\label{tab:supp_ablation_dhp}
\end{table*}

Removing MST-based global decoding or cross-page edges yields the largest degradations, confirming that globally valid tree decoding and long-range document links are the two most important ingredients for stable hierarchy recovery. SoftROI, header-centric parent priors, and candidate pruning provide smaller but still consistent gains.

\subsection{Per-corpus and Per-type Breakdowns}

The main paper reports macro-averaged DHP, retrieval, and QA metrics across datasets.
Here we additionally break down DHP performance by edge type and analyze how different
methods behave on structurally difficult subsets.

\paragraph{Per-type DHP analysis.}

% ===== DHP per-type F1 (DocHieNet + HRDH + HRDS macro-average) =====

% ----- General LVLM (GPT-5, LLaVA, InternVL, Qwen2.5-VL average) -----
\newcommand{\DHPGenLocal}{30.4}
\newcommand{\DHPGenCross}{24.1}
\newcommand{\DHPGenFig}{4.3}

% ----- DHP-Average (DocParser, DSG, DSPS, DSHP-LLM average) -----
\newcommand{\DHPAvgLocal}{57.2}
\newcommand{\DHPAvgCross}{46.5}
\newcommand{\DHPAvgFig}{8.9}

% ----- Qwen2.5-DHP-SFT (tree-aware + LVLM fine-tuning) -----
\newcommand{\DHPQwenLocal}{58.6}
\newcommand{\DHPQwenCross}{47.9}
\newcommand{\DHPQwenFig}{32.4}

% ----- M3DocDep (ours) -----
\newcommand{\DHPMthreeLocal}{88.1}
\newcommand{\DHPMthreeCross}{79.2}
\newcommand{\DHPMthreeFig}{69.3}
% ============================================================================

\begin{figure}[t]
\centering
\begin{tikzpicture}
\begin{axis}[
  ybar,
  bar width=5pt,
  width=0.39\textwidth,
  height=3.7cm,
  ymin=0, ymax=100,
  enlarge x limits=0.20,
  symbolic x coords={Local,Cross-page,Fig./Table},
  xtick=data,
  xticklabel style={font=\scriptsize, rotate=15, anchor=east},
  yticklabel style={font=\scriptsize},
  ylabel={F1 (\%)},
  ylabel style={font=\scriptsize},
  title style={font=\footnotesize},
  legend style={
    font=\scriptsize,
    at={(0.5,1.16)},
    anchor=south,
    legend columns=4
  },
]

% ---- General LVLM ----
\addplot coordinates {
  (Local,\DHPGenLocal)
  (Cross-page,\DHPGenCross)
  (Fig./Table,\DHPGenFig)
};

% ---- DHP-Average ----
\addplot coordinates {
  (Local,\DHPAvgLocal)
  (Cross-page,\DHPAvgCross)
  (Fig./Table,\DHPAvgFig)
};

% ---- Qwen2.5-DHP-SFT ----
\addplot coordinates {
  (Local,\DHPQwenLocal)
  (Cross-page,\DHPQwenCross)
  (Fig./Table,\DHPQwenFig)
};

% ---- M3DocDep (ours) ----
\addplot coordinates {
  (Local,\DHPMthreeLocal)
  (Cross-page,\DHPMthreeCross)
  (Fig./Table,\DHPMthreeFig)
};

\legend{General LVLM, DHP-Average, Qwen2.5-DHP-SFT, M3DocDep}

\end{axis}
\end{tikzpicture}
\caption{Per-type DHP performance over Local, Cross-page, and Fig./Table edge subsets for General LVLM, DHP-Average, Qwen2.5-DHP-SFT, and M3DocDep, macro-averaged over DocHieNet, HRDH, and HRDS.}

\label{fig:per_type_dhp}
\end{figure}
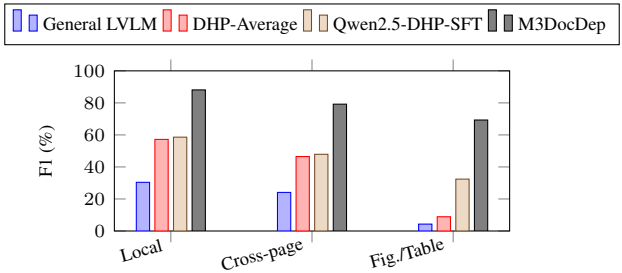

Figure~\ref{fig:per_type_dhp} reports parent-prediction F1(\%) on three edge subsets:
\emph{Local} (child and parent on the same page), \emph{Cross-page} (child and
parent on different pages), and \emph{Fig./Table} (child block type is
\texttt{figure} or \texttt{table}), macro-averaged over DocHieNet, HRDH, and HRDS.
Across all methods the Local subset is always easier than Cross-page and Fig./Table,
and General LVLMs as well as classical DHP parsers show only moderate accuracy on
Local edges while almost completely failing to recover figure/table relations.
Qwen2.5-DHP-SFT partially closes this gap, especially for Fig./Table edges, but still
lags behind on cross-page structure.
In contrast, M3DocDep clearly dominates on all three subsets in the plot:
it maintains strong accuracy not only on Local edges but also on cross-page links,
and is the only setting that achieves high, stable performance on Fig./Table edges. This suggests that M3DocDep is particularly
effective at capturing long-range structure and image/table-centric regions that other approaches largely miss.

\subsection{Sensitivity to the Number of Retrieved Chunks}

\begin{figure}[t]
\centering
\begin{tikzpicture}
\begin{axis}[
  width=0.42\textwidth,
  height=3.5cm,
  xmin=1, xmax=4,
  ymin=0.24, ymax=0.60,
  xtick={1,2,3,4},
  xticklabel style={font=\scriptsize},
  yticklabel style={font=\scriptsize},
  xlabel={$k$},
  ylabel={Macro-avg.\ nDCG@{$k$}},
  xlabel style={font=\scriptsize},
  ylabel style={font=\scriptsize},
  legend style={
    font=\scriptsize,
    at={(0.5,1.15)},
    anchor=south,
    legend columns=3
  },
  grid=both,
  grid style={dotted,gray!40},
]

% ---- Semantic (macro-avg) ----
\addplot[
  color=gray!60,
  mark=*,
  mark size=1.4pt,
  dashed,
  thick,
] coordinates {
  (1,0.2509) (2,0.2721) (3,0.2879) (4,0.2980)
};

% ---- Structure-based (macro-avg) ----
\addplot[
  color=gray!50,
  mark=triangle*,
  mark size=1.6pt,
  dashed,
  thick,
] coordinates {
  (1,0.3904) (2,0.4245) (3,0.4427) (4,0.4542)
};

% ---- Perplexity (macro-avg) ----
\addplot[
  color=gray!65,
  mark=square*,
  mark size=1.5pt,
  dashed,
  thick,
] coordinates {
  (1,0.4107) (2,0.4489) (3,0.4660) (4,0.4780)
};

% ---- LumberChunker (macro-avg) ----
\addplot[
  color=gray!70,
  mark=diamond*,
  mark size=1.8pt,
  dashed,
  thick,
] coordinates {
  (1,0.4113) (2,0.4506) (3,0.4675) (4,0.4793)
};

% ---- Length (macro-avg) ----
\addplot[
  color=black,
  mark=o,
  mark size=1.6pt,
  thick,
] coordinates {
  (1,0.4413) (2,0.4727) (3,0.4911) (4,0.5033)
};

% ---- MultiDocFusion (macro-avg) ----
\addplot[
  color=orange!80!black,
  mark=square*,
  mark size=1.8pt,
  very thick,
] coordinates {
  (1,0.4644) (2,0.4984) (3,0.5147) (4,0.5246)
};

% ---- M3DocDep (macro-avg) ----
\addplot[
  color=cyan!70!blue,
  mark=*,
  mark size=2.0pt,
  ultra thick,
] coordinates {
  (1,0.5021) (2,0.5400) (3,0.5583) (4,0.5695)
};

\legend{Semantic, Structure-based, Perplexity, LumberChunker, Length, MultiDocFusion, M3DocDep}

\end{axis}
\end{tikzpicture}
\caption{Macro-averaged nDCG@$k$ over DUDE, MP-DocVQA, CUAD, and MOAMOB
for $k\in\{1,2,3,4\}$.
\textbf{M3DocDep} dominates across all $k$, with the largest margins at small $k$,
and remains consistently ahead as $k$ grows.}
\label{fig:kret_macro_avg}
\end{figure}
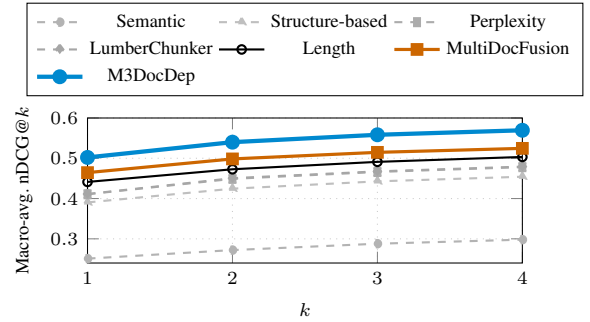

We assess robustness to the retrieval budget by plotting the \emph{macro-averaged}
nDCG@$k$ over DUDE, MP-DocVQA, CUAD, and MOAMOB for $k\in\{1,2,3,4\}$
(Figure~\ref{fig:kret_macro_avg}). Across all $k$, \textbf{M3DocDep} consistently yields the
best nDCG and preserves a clear margin over every baseline; the advantage is
most pronounced under tight budgets (small $k$) and remains visible as $k$
grows. These results indicate that tree-guided, structure-aware chunking
provides high-quality evidence with few retrieved chunks and scales gracefully
to larger retrieval budgets.

\subsection{Chunk Length and Distribution}

By design, our tree-guided chunking keeps chunk lengths
within a moderate target range while respecting section
boundaries, avoiding both tiny fragments and excessively
large chunks. In contrast, purely length-based or semantic
chunkers often produce a mix of very short and very long
chunks. This design helps dense retrievers by aligning
chunks more closely with underlying semantic units.

% ============================================================

\section{Qualitative Case Studies and Error Analysis}
\label{sec:qualitative_supp}

\subsection{Chunking Comparisons}

\begin{figure*}[t]
\centering
\resizebox{\linewidth}{!}{%
\begin{tikzpicture}[font=\scriptsize, x=1cm,y=1cm]

  % Common styles
  \tikzset{
    chunkframe/.style={%
      rounded corners,
      draw=blue!70!black,
      thick,
      inner sep=6pt
    },
    chunksec/.style={%
      fill=blue!10,
      text width=3.5cm,
      align=left,
      inner sep=3pt,
      font=\scriptsize\ttfamily
    },
    chunkmeta/.style={%
      fill=yellow!10,
      text width=3.2cm,
      align=left,
      inner sep=3pt
    },
    chunkctx/.style={%
      fill=green!8,
      text width=3.2cm,
      align=left,
      inner sep=3pt
    }
  }

  % ======================================================
  % (a) Text-based Chunking
  % ======================================================
  \node[chunkctx,anchor=north west] (actx)
    at (0,0) {%
      Light deflection in curved spacetimes is one of the earliest predictions of Einstein's general theory of relativity \dots\ ICOb $b_c$ $r_c$ $\varphi$ $r\to\infty$ $\varphi\to\alpha$ $\varphi\to -\pi$ XY $\varphi_0$ Photon Trajectory Critical Photon Trajectory Black Hole Figure~1: Schematic of photon trajectory in the equatorial plane of a Kerr black hole, parametrized as $r = r(\varphi)$. The $+Z$ direction is out of the page \dots\ In our previous paper~[6] we examined the deflection of photons traveling in the equatorial plane of a Kerr (spinning) black hole \dots
    };
  
  \node[chunkframe,fit=(actx),inner sep=10pt] (a) {};
  
  \node[anchor=south,font=\scriptsize\bfseries]
    at ([yshift=3pt]a.north) {(a) Text-based chunking};

  % ======================================================
  % (b) Structure-based Chunking
  % ======================================================
  \node[chunkctx,anchor=north west] (bctx)
    at (4.8,0) {%
      [Figure] ICOb $b_c$ $r_c$ $\varphi$ $r\to\infty$ $\varphi\to\alpha$ $\varphi\to -\pi$ XY $\varphi_0$ Photon Trajectory Critical Photon Trajectory Black Hole\\[2pt]
      Figure~1: Schematic of photon trajectory in the equatorial plane of a Kerr black hole, parametrized as $r = r(\varphi)$. The labeled photon trajectory shows the relationship between the impact parameter $b$, radial distance $r$, azimuthal angle $\varphi$, and bending angle $\alpha$ \dots
    };
  \node[chunkframe,fit=(bctx),inner sep=10pt] (b) {};
  \node[anchor=south,font=\scriptsize\bfseries]
    at ([yshift=3pt]b.north) {(b) Structure-based chunking};

  % ======================================================
  % (c) MultiDocFusion Chunking  -- section_path + context
  % ======================================================
  \node[chunksec,anchor=north west] (csec)
    at (9.6,0) {%
      \#\  |  Accurate closed-form trajectories of light around a Kerr black hole using asymptotic approximants  \newline
\#\# 2.~Light deflection: notation and conventions  \newline
    };
  \node[chunkctx,anchor=north west] (cctx)
    at ([yshift=-0.15cm]csec.south west) {%
      ICOb $b_c$ $r_c$ $\varphi$ $r\to\infty$ $\varphi\to\alpha$ $\varphi\to -\pi$ XY $\varphi_0$ Photon Trajectory Critical Photon Trajectory Black Hole\\[2pt]
      Figure~1: Schematic of photon trajectory in the equatorial plane of a Kerr black hole, parametrized as $r = r(\varphi)$. The $+Z$ direction is out of the page and the labeled photon trajectory shows the relationship between $b$, $r$, $\varphi$, and $\alpha$ \dots
    };
  
  \node[chunkframe,fit=(csec)(cctx),inner sep=10pt] (c) {};
  \node[anchor=south,font=\scriptsize\bfseries]
    at ([yshift=3pt]c.north) {(c) MultiDocFusion chunking};

  % ======================================================
  % (d) M3DocDep Chunking  -- section_path + tree + PNG + caption
  % ======================================================
  \node[chunksec,anchor=north west] (dsec)
    at (14.4,0) {%
      \#\  |  Accurate closed-form trajectories of light around a Kerr black hole using asymptotic approximants  \newline
\#\# 1.~Introduction  \newline
    };
  \node[chunkmeta,anchor=north west] (dmeta)
    at ([yshift=-0.12cm]dsec.south west) {%
      page : 2
    };

  \node[anchor=north west,inner sep=0pt] (dimg)
    at ([yshift=-0.18cm,xshift=0.15cm]dmeta.south west) {%
      \includegraphics[width=3.2cm]{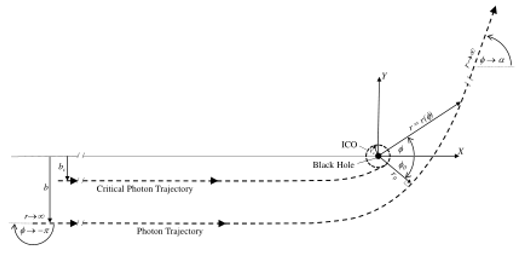}
      % \fbox{\parbox[c][1.6cm][c]{3.2cm}{\centering PNG block}}
    };

  \node[chunkctx,anchor=north west,text width=3.2cm] (dctx)
    at ([yshift=-0.15cm]dimg.south west) {%
      Figure~1: Schematic of photon trajectory in the equatorial plane of a Kerr black hole, parametrized as $r = r(\varphi)$. The labeled photon trajectory shows the relationship between the impact parameter $b$, radial distance $r$, azimuthal angle $\varphi$, and bending angle $\alpha$ \dots
    };

  \node[chunkframe,fit=(dsec)(dmeta)(dimg)(dctx),inner sep=10pt] (d) {};
  \node[anchor=south,font=\scriptsize\bfseries]
    at ([yshift=3pt]d.north) {(d) M3DocDep chunking};

\end{tikzpicture}%
}

\caption{Comparison of four chunking paradigms on the same figure region: text-based chunking mixes surrounding text and OCR'ed figure text; structure-based chunking isolates figure and caption as text only; MultiDocFusion adds a section\_path over text; M3DocDep attaches the correct section path, preserves the shared metadata fields when available, and keeps the figure region aligned with its caption in the same chunk representation.}
\label{fig:chunking_figure_region}
\end{figure*}

Using representative documents from VQA Datasets, we overlay chunk boundaries from different methods (Length, Semantic, LumberChunker, Perplexity, Structure-based, MultiDocFusion) directly on page images. Color-coding shows that text-based methods often split sections mid-paragraph or separate figures from captions, whereas M3DocDep aligns chunk boundaries with section subtrees and keeps visual content with its description. More details of the chunking method examples are shown in Tab.~\ref{tab:chunking_examples}. Figures~\ref{fig:document_example_steps_cd_real} and~\ref{fig:chunking_figure_region} make the MST-decoded tree, the induced chunks, and their multimodal bindings concrete; together they visualize the exact tree-to-chunk pathway used in the main method.

\paragraph{Handling of figure/image regions.}
Figure~\ref{fig:chunking_figure_region} zooms in on a single
figure region and compares how
different chunking strategies represent the same content.
\emph{Text-based chunking} operates purely on flattened OCR
text, so the figure text is mixed with surrounding paragraphs and
no explicit notion of a figure region is preserved. \emph{Structure-based
chunking} removes this entanglement by isolating the figure and
its caption, but still treats them as plain text only.
\emph{MultiDocFusion} augments the structure-based chunk with
an LLM-predicted \texttt{section\_path}, yet this path can be
misaligned with the true document hierarchy because it is inferred
from text alone. In contrast, \emph{M3DocDep} attaches the correct
\texttt{section\_path} from the global dependency tree and keeps the
figure region aligned with its textual caption inside the same chunk
representation, so figure--caption context is preserved rather than
being split across unrelated text. Consequently, M3DocDep is the
only compared chunking method that consistently preserves
figure/image regions as first-class multimodal units, rather than
collapsing them into text-only representations.

\subsection{Failure Cases}

We qualitatively observe the following typical failure modes:

\begin{itemize}[leftmargin=*]
  \item \textbf{parser misses and block fragmentation:} upstream detectors may split a logical region into several small blocks or merge nearby regions, which destabilizes the candidate-parent set before tree decoding;
  \item \textbf{OCR corruption in degraded scans:} missing or heavily garbled text weakens both the block embedding and the serialized chunk, especially for densely scanned manuals and contracts;
  \item \textbf{ambiguous or implicit headings:} some documents signal section transitions only through typography or whitespace, making header-centric parent selection harder;
  \item \textbf{repeated headers/footers and template artifacts:} boilerplate repeated across pages can attract spurious parents if the visual hierarchy is weak;
  \item \textbf{caption drift and multi-page figures/tables:} long visual regions that span pages or sit far from their captions can still be attached incorrectly when neither tree evidence nor spatial fallback is strong enough.
\end{itemize}

In such cases the dependency head may attach blocks to suboptimal parents or fail to link the correct captions, leading to imperfect trees and suboptimal chunks. These failures are nevertheless informative: they show that the remaining bottlenecks are concentrated in upstream block quality, OCR corruption, and highly ambiguous layouts rather than in ordinary section-level documents. We consider joint training with layout-normalized variants, stronger weak supervision, and query-aware reranking as promising directions.

% ============================================================

\section{Runtime and Scalability Analysis}
\label{sec:runtime_supp}

\subsection{Runtime Breakdown}

\begin{table}[h]
\centering
\scriptsize
\setlength{\tabcolsep}{4pt}
\renewcommand{\arraystretch}{1.08}
\begin{tabularx}{\columnwidth}{@{}>{\raggedright\arraybackslash}Xc@{}}
\toprule
\textbf{Component} & \textbf{sec/page or memory} \\
\midrule
SharedDet (DP+OCR) & 3.2 sec/page \\
Core (LVLM + SoftROI + scoring + MST) & 0.4 sec/page \\
Total end-to-end & 3.6 sec/page \\
Core throughput & 2.5 pages/s \\
Peak GPU memory & 27 GB \\
LVLM-only autoregressive hierarchy generation & 20 sec/page \\
\bottomrule
\end{tabularx}
\caption{Runtime summary on a single A100 (80GB). The lightweight M3DocDep core is substantially faster than LVLM-only autoregressive hierarchy generation.}
\label{tab:supp_runtime}
\end{table}

We measure wall-clock runtime on a single A100 (80GB) and separate the end-to-end indexing cost into SharedDet and the M3DocDep core. The dominant cost is upstream DP+OCR (3.2 sec/page), while the core module---LVLM forward, SoftROI pooling, edge scoring, and MST decoding---requires only 0.4 sec/page and reaches 2.5 pages/s at 27 GB peak GPU memory. The runtime breakdown is shown in Table~\ref{tab:supp_runtime}.

\subsection{Scaling with Document Length}

Because DP and OCR operate per page and dependency
scoring is restricted to a small parent candidate set
($K \ll N$), the effective complexity of M3DocDep
is near-linear in the number of blocks/pages for typical
industrial documents. Empirically, the time per document
grows roughly linearly with page count, and larger docs can
be batched or parallelized. In practice, the core M3DocDep module contributes only a small fraction of the total latency, so larger deployments can parallelize SharedDet aggressively while keeping the dependency and chunking stage lightweight.

% ============================================================

\section{Limitations and Broader Impact}
\label{sec:limits}

\paragraph{Limitations.}
M3DocDep relies on a frozen upstream DP detector and OCR engine; when these components produce fragmented or merged blocks on heavily degraded scans, the downstream dependency head inherits these errors, as discussed in the failure cases (Sec.~\ref{sec:qualitative_supp}). The biaffine scoring head is trained on three DHP corpora that, while diverse, do not cover all industrial document types (e.g., handwritten forms, non-Latin scripts beyond Korean and Chinese). The pipeline currently constructs dependency trees per document; extending it to model inter-document relations (e.g., cross-references between contract annexes) remains future work. Finally, inference speed is dominated by the frozen LVLM forward pass and SharedDet, which may limit deployment on very large corpora without further engineering (e.g., distillation, quantization).

\paragraph{Broader impact.}
By improving the accuracy of document structure recovery and chunk construction, M3DocDep can help users retrieve more reliable answers from long, complex documents, which has positive implications for domains such as legal review, technical maintenance, and financial auditing. We do not foresee direct negative societal impacts specific to our method; however, as with any RAG system, downstream answer quality depends on the accuracy of the source documents, and users should verify critical information independently.

% --------- Table 1/2: Length ~ Perplexity ---------
\begin{table*}[t]
  \centering
  \scriptsize
  \renewcommand{\arraystretch}{1.1}
  \setlength{\tabcolsep}{4pt}
  \begin{tabular}{@{}l l p{0.7\linewidth}@{}}
    \toprule
    \textbf{Method} & \textbf{Chunk} & \textbf{Example Content} \\
    \midrule
    \multirow{3}{*}{Length chunking}
      & Chunk 1 & Accurate closed-form trajectories of light around a Kerr black hole using asymptotic approximants --- Ryne J.~Beachley1, Morgan Mistysyn2, Joshua A.~Faber1,4, Steven J.~Weinstein3,4, Nathaniel S.~Barlow1,4. 1 School of Mathematical Sciences, Rochester Institute of Technology, Rochester, NY 14623 \dots\ Abstract. Highly accurate closed-form expressions that describe the full trajectory of photons propagating in the equatorial plane of a Kerr black hole are obtained using asymptotic approximants \dots \\ \addlinespace[3pt]
      & Chunk 2 & This work extends a prior study of the overall bending angle for photons (Barlow et al.\ 2017, Class. Quantum Grav., 34, 135017). The expressions obtained provide accurate trajectory predictions for arbitrary spin and impact parameters, and provide significant time advantages compared with numerical evaluation of the elliptic integrals that describe photon trajectories \dots\ Keywords: Geodesics, Light deflection, Kerr black holes, Asymptotic approximants. Submitted to: Class. Quantum Grav. \dots \\ \addlinespace[3pt]
      & Chunk 3 & 1. Introduction Light deflection in curved spacetimes is one of the earliest predictions of Einstein's general theory of relativity, and one of the best understood aspects of the theory. The null geodesics describing photon trajectories have been investigated for a wide variety of physical configurations in a number of limits \dots\ After the initial construction of the Kerr metric describing spinning black holes, many of the early results on null geodesics in these spacetimes were derived by Carter \dots \\ 
    \midrule
    \multirow{3}{*}{Semantic chunking}
      & Chunk 1 & Accurate closed-form trajectories of light around a Kerr black hole using asymptotic approximants --- Ryne J.~Beachley1, Morgan Mistysyn2, Joshua A.~Faber1,4, Steven J.~Weinstein3,4, Nathaniel S.~Barlow1,4. 1 School of Mathematical Sciences, Rochester Institute of Technology, Rochester, NY 14623; 2 Department of Industrial and Systems Engineering; 3 Department of Chemical Engineering; 4 Center for Computational Relativity and Gravitation, Rochester Institute of Technology \dots\ E-mail: nsbsma@rit.edu \dots \\ \addlinespace[3pt]
      & Chunk 2 & Abstract. Highly accurate closed-form expressions that describe the full trajectory of photons propagating in the equatorial plane of a Kerr black hole are obtained using asymptotic approximants. This work extends a prior study of the overall bending angle for photons \dots\ To construct approximants, asymptotic expansions for photon deflection are required in various limits \dots\ new coefficients are reported for the bending angle in the weak-field limit (large impact parameter) \dots \\ \addlinespace[3pt]
      & Chunk 3 & 1. Introduction Light deflection in curved spacetimes is one of the earliest predictions of Einstein's general theory of relativity \dots\ The limit where photons approach the innermost circular orbit (ICO), referred to as the strong-field limit, has also been explored for decades \dots\ We refer readers to Chandrasekhar's work on the subject for a thorough review on geodesics in black hole spacetimes \dots \\
    \midrule
    \multirow{3}{*}{LumberChunker}
      & Chunk 1 & Accurate closed-form trajectories of light around a Kerr black hole using asymptotic approximants --- [title and front matter] Ryne J.~Beachley1, Morgan Mistysyn2, Joshua A.~Faber1,4, Steven J.~Weinstein3,4, Nathaniel S.~Barlow1,4 \dots\ Abstract. Highly accurate closed-form expressions that describe the full trajectory of photons propagating in the equatorial plane of a Kerr black hole are obtained using asymptotic approximants \dots \\ \addlinespace[3pt]
      & Chunk 2 & Abstract. Highly accurate closed-form expressions that describe the full trajectory of photons \dots\ The expressions obtained provide accurate trajectory predictions for arbitrary spin and impact parameters \dots\ Keywords: Geodesics, Light deflection, Kerr black holes, Asymptotic approximants. Submitted to: Class. Quantum Grav. 1. Introduction Light deflection in curved spacetimes is one of the earliest predictions of Einstein's general theory of relativity \dots \\ \addlinespace[3pt]
      & Chunk 3 & 1. Introduction Light deflection in curved spacetimes is one of the earliest predictions of Einstein's general theory of relativity \dots\ The limit where photons approach the innermost circular orbit (ICO) has also been explored for decades \dots\ Figure 1: Schematic of photon trajectory in the equatorial plane of a Kerr black hole, parametrized as $r = r(\varphi)$, showing the relationship between impact parameter $b$, radial distance $r$, azimuthal angle $\varphi$, and bending angle $\alpha$ \dots \\
    \midrule
    \multirow{3}{*}{Perplexity chunking}
      & Chunk 1 & Accurate closed-form trajectories of light around a Kerr black hole using asymptotic approximants --- Ryne J.~Beachley1, Morgan Mistysyn2, Joshua A.~Faber1,4, Steven J.~Weinstein3,4, Nathaniel S.~Barlow1,4 \dots\ 1 School of Mathematical Sciences, Rochester Institute of Technology, Rochester, NY 14623 \dots\ Keywords: Geodesics, Light deflection, Kerr black holes, Asymptotic approximants \dots \\ \addlinespace[3pt]
      & Chunk 2 & Abstract. Highly accurate closed-form expressions that describe the full trajectory of photons propagating in the equatorial plane of a Kerr black hole are obtained using asymptotic approximants. This work extends a prior study of the overall bending angle for photons (Barlow et al.\ 2017, Class. Quantum Grav., 34, 135017). The expressions obtained provide accurate trajectory predictions for arbitrary spin and impact parameters \dots \\ \addlinespace[3pt]
      & Chunk 3 & Light deflection in curved spacetimes is one of the earliest predictions of Einstein's general theory of relativity, and one of the best understood aspects of the theory \dots\ The limit where photons approach the innermost circular orbit (ICO) has also been explored for decades \dots\ Figure 1: Schematic of photon trajectory in the equatorial plane of a Kerr black hole, parametrized as $r = r(\varphi)$ \dots \\
    \bottomrule
  \end{tabular}
  \caption{Qualitative comparison of chunking methods applied to the document in Fig.~\ref{fig:document_example_steps_ab_real} (part 1/2).}
  \label{tab:chunking_examples}
\end{table*}

% --------- Table 2/2: Structure-based ~ M3DocDep ---------

\begin{table*}[t]
  \centering
  \scriptsize
  \renewcommand{\arraystretch}{1.1}
  \setlength{\tabcolsep}{4pt}
  \begin{tabular}{@{}l l p{0.7\linewidth}@{}}
    \toprule
    \textbf{Method} & \textbf{Chunk} & \textbf{Example Content} \\
    \midrule
    \multirow{3}{*}{Structure-based}
      & Chunk 1 & [Title block] Accurate closed-form trajectories of light around a Kerr black hole using asymptotic approximants --- Ryne J.~Beachley1, Morgan Mistysyn2, Joshua A.~Faber1,4 \dots\ Abstract. Highly accurate closed-form expressions that describe the full trajectory of photons \dots\ Keywords: Geodesics, Light deflection, Kerr black holes. Submitted to: Class. Quantum Grav. \\ \addlinespace[3pt]
      & Chunk 2 & [Section: 1.\ Introduction] Light deflection in curved spacetimes is one of the earliest predictions of Einstein's general theory of relativity \dots\ The limit where photons approach the innermost circular orbit (ICO) has also been explored for decades \dots \\ \addlinespace[3pt]
      & Chunk 3 & [Figure + Caption] Figure~1: Schematic of photon trajectory in the equatorial plane of a Kerr black hole, parametrized as $r = r(\varphi)$. The labeled photon trajectory shows the relationship between the impact parameter $b$, radial distance $r$, azimuthal angle $\varphi$, and bending angle $\alpha$ \dots \\
    \midrule
    \multirow{3}{*}{MultiDocFusion}
      & Chunk 1 & \texttt{section\_path:} \# Accurate closed-form trajectories \dots \newline [Title and front matter] Ryne J.~Beachley1 \dots\ Abstract. Highly accurate closed-form expressions \dots\ Keywords: Geodesics, Light deflection, Kerr black holes \dots \\ \addlinespace[3pt]
      & Chunk 2 & \texttt{section\_path:} \# Accurate closed-form trajectories $>$ \#\# 1.\ Introduction \newline Light deflection in curved spacetimes is one of the earliest predictions of Einstein's general theory of relativity \dots\ The null geodesics describing photon trajectories have been investigated for a wide variety of physical configurations \dots \\ \addlinespace[3pt]
      & Chunk 3 & \texttt{section\_path:} \# Accurate closed-form trajectories $>$ \#\# 2.\ Light deflection \newline Figure~1: Schematic of photon trajectory in the equatorial plane of a Kerr black hole, parametrized as $r = r(\varphi)$ \dots \\
    \midrule
    \multirow{3}{*}{\textbf{M3DocDep}}
      & Chunk 1 & \texttt{section\_path:} \# Accurate closed-form trajectories \dots \newline \texttt{pages:} 1 \newline [Title and front matter] Ryne J.~Beachley1 \dots\ Abstract. Highly accurate closed-form expressions \dots\ Keywords: Geodesics, Light deflection, Kerr black holes. Submitted to: Class. Quantum Grav. \\ \addlinespace[3pt]
      & Chunk 2 & \texttt{section\_path:} \# Accurate closed-form trajectories $>$ \#\# 1.\ Introduction \newline \texttt{pages:} 1--2 (cross-page) \newline Light deflection in curved spacetimes is one of the earliest predictions of Einstein's general theory of relativity \dots\ The null geodesics describing photon trajectories have been investigated for a wide variety of physical configurations in a number of limits \dots \\ \addlinespace[3pt]
      & Chunk 3 & \texttt{section\_path:} \# Accurate closed-form trajectories $>$ \#\# 1.\ Introduction \newline \texttt{pages:} 2 \newline [Figure region + Caption] Figure~1: Schematic of photon trajectory in the equatorial plane of a Kerr black hole, parametrized as $r = r(\varphi)$. The labeled photon trajectory shows the relationship between the impact parameter $b$, radial distance $r$, azimuthal angle $\varphi$, and bending angle $\alpha$ \dots \\
    \bottomrule
  \end{tabular}
  \caption{Qualitative comparison of chunking methods applied to the document in Fig.~\ref{fig:document_example_steps_ab_real} (part 2/2). Structure-based chunking isolates visual regions; MultiDocFusion adds section paths; \textbf{M3DocDep} additionally preserves page metadata and keeps figure regions with captions.}
  \label{tab:chunking_examples_part2}
\end{table*}

\FloatBarrier

\clearpage

% End of Supplement

\end{document}